\documentclass[useAMS,usenatbib]{mn2e}
\usepackage{amsmath}                                                            
\usepackage{amsfonts}
\usepackage{graphicx}
\usepackage{float}
\usepackage{morefloats}
\usepackage{listings}

\title[PRIMAL: A PaRtIcle-by-particle M2M ALgorithm]
{Investigating Bar Structure of Disc Galaxies via PRIMAL: A PaRtIcle-by-particle M2M ALgorithm}
\author[J. A. S. Hunt, D. Kawata \& H. Martel]
 {Jason A. S. Hunt,$^{1}$\thanks{E-mail: jash2@mssl.ucl.ac.uk}
Daisuke Kawata$^{1}$ and
Hugo Martel$^{2,3}$
\\
$^{1}$ Mullard Space Science Laboratory, University College London,
Holmbury St. Mary, Dorking, Surrey, RH5 6NT, UK \\
$^2$ D\'{e}partment de physique, de g\'{e}nie physique et d'optique, Universit\'{e} Laval, Qu\'{e}bec, QC, Canada
\\
$^3$ Centre de Recherche en Astrophysique du Qu\'{e}bec, C.P. 6128, Succ. Centre-Ville, Montr\'{e}al, QC, Canada.  
\\
}
\date{Accepted 2013 April 16.  Received 2013 April 8; in original form 2013 February 4} 

\pagerange{\pageref{firstpage}--\pageref{lastpage}}
\pubyear{2012}

\begin{document}

\maketitle

\label{firstpage}

\begin{abstract}
We have modified our particle-by-particle adaptation of the made-to-measure (M2M) method, with the aim of modelling the Galactic disc from upcoming Galactic stellar survey data. In our new particle-by-particle M2M algorithm, \sc{primal}\rm, the observables of the target system are compared with those of the model galaxy at the position of the target stars, i.e. particles. The mass of the model particles are adjusted to reproduce the observables of the target system, and the gravitational potential is automatically adjusted by the changing mass of the particles. This paper builds upon our previous work, introducing likelihood-based velocity constraints in \sc{primal}\rm. In this paper we apply \sc{primal }\rm to barred disc galaxies created by a $N$-body simulation in a known dark matter potential, with no error in the observables. This paper demonstrates that \sc{primal }\rm can recover the radial profiles of the surface density, velocity dispersion in the radial and perpendicular directions, and the rotational velocity of the target discs, along with the apparent bar structure and pattern speed of the bar, especially when the reference frame is adjusted so that the bar angle of the target galaxy is aligned to that of the model galaxy at every timestep. 
\end{abstract}


\section{Introduction}
\label{intro-sec}
The made-to-measure (M2M) method developed by \cite{ST96} has seen increasing use in the past few years, including multiple papers involving \sc{nmagic }\rm \citep[e.g.][]{DeL07,DeL08,DGMT11,MG12}, and the series of three papers \citep{LM10,LM12,LMIII} based on the M2M algorithm described in \cite{LM10}. This increasing level of interest is now highlighting the potential of the M2M method to be used in many different applications, from probing the dark matter halo of NGC 4649 \citep{DGMT11} to tailoring $N$-body initial conditions \citep{Deh09}. While previous work has primarily focused on theoretical models and external galaxies, \cite{BDG04} and \cite{LMIII} have applied their versions of M2M to the Milky Way which is also our eventual goal. 

The Milky Way is known to be a barred spiral \citep{SB90,W92,BGS97} but there is still disagreement over whether the bar is formed of a single structure, or if it is comprised of a separate long flat bar in addition to the barred bulge \citep[e.g.][]{MG11,A12,VVV12}. The bar angle, defined as an offset between the axis of the bar and the Sun-Galactic centre line is also still debated. For example \cite{Detal95} suggested a bar angle of $20^{\text{o}}\pm10^{\text{o}}$, \cite{Aetal00} found $15^{\text{o}}$, \cite{RC08} found $20^{\text{o}}-35^{\text{o}}$, \cite{MNQ11} found $20^{\text{o}}-45^{\text{o}}$ and \cite{Getal11} found $45^{\text{o}}$. There is also still discussion of the pattern speed of the bar, $\Omega_p$. Among others \cite{F99} calculated a value of $\Omega_p\approx50 \text{ km s}^{-1}\text{kpc}^{-1}$, \cite{D99} calculated $\Omega_p=53\pm3 \text{ km s}^{-1}\text{kpc}^{-1}$, \cite{RC08} calculated $\Omega_p=30-40 \text{ km s}^{-1}\text{kpc}^{-1}$, \cite{WZMR12} concluded $\Omega_p\approx60 \text{ km s}^{-1}\text{kpc}^{-1}$ and \cite{G11} provided a review of previous work, concluding a likely pattern speed for the bar of $\Omega_p\approx50-60 \text{ km s}^{-1}\text{kpc}^{-1}$.


\cite{BDG04} were the first to apply the original M2M algorithm from \cite{ST96} to the Milky Way, which was an important first test of M2M on the Milky Way, matching observed bar kinematics in several fields. The M2M algorithm has undergone significant improvements since then much more observational data have become available, leaving the door open for more up to date M2M modelling of the Milky Way. 

\cite{LMIII} have taken the next step, applying their M2M algorithm from \cite{LM10} to observed radial velocity data from the Bulge RAdial Velocity Assay (BRAVA)\footnote{http://brava.astro.ucla.edu/data.htm} \citep{RRHZ07,Ketal12} and the particle mass distribution of an $N$-body barred galaxy model \citep{SRKHPK10} that has been built to reproduce the Milky Way disc and earlier BRAVA data. \cite{LMIII} used the gravitational potential calculated from the $N$-body model of \cite{SRKHPK10}. Then, they rotate this potential with a fixed pattern speed and assumed bar angle from the line of the Galactic centre and the observer, i.e. the position of the Sun. They ran many models with different pattern speeds and bar angles and explored the models which best fit their observables. They found their best model recovered the bar angle and pattern speed of the \cite{SRKHPK10} $N$-body model, and reproduced the mean radial velocity and radial velocity dispersion of the BRAVA data very well.

Although \cite{LMIII} have taken an important step forwards there will be much room for improvement for applying M2M to the future observational data. The European Space Agency's upcoming Gaia mission along with ground based surveys, e.g. the Panoramic Survey Telescope And Rapid Response System \citep[PanStarrs, e.g.][]{Ketal10}, the Visible and Infra-red Survey Telescope for Astronomy \citep[VISTA, e.g.][]{Metal09}, the Large Synoptic Survey Telescope \citep[LSST, e.g.][]{Ietal08}, the Sloan Extension for Galactic Understanding and Exploration \citep[SEGUE, e.g.][]{Yea09}, the Apache Point Observatory Galactic Evolution Experiment \citep[APOGEE, e.g.][]{APetal08}, Subaru-PFS \citep[e.g][]{Eetal12-2}, The Radial Velocity Experiment \citep[RAVE, e.g.][]{SEA06}, the Large sky Area Multi-Object fibre Spectroscopic Telescope, LAMOST Experiment for Galactic Understanding and Explanation \citep[LEGUE, e.g.][]{Detal12} and the Gaia-ESO public spectrographic survey \citep[e.g.][]{Metal12} will provide us with an unprecedentedly large amount of data about the Milky Way, which we can use as observational constraints for our M2M algorithm. As such we have made modifications to our algorithm with this goal in mind.

In \cite{HK12}, hereafter Paper 1, we developed a particle-by-particle M2M algorithm now called \sc{primal }\rm (PaRtIcle-by-particle M2M ALgorithm). Because the Galactic stellar-survey data, such as the ones Gaia will produce, are in the form of the position and velocity of individual stars, \sc{primal }\rm is designed to compare the observables at the position of each star, i.e. not binned data as in previous M2M modelling. Another major difference between\sc{primal }\rm and other M2M algorithms is that the gravitational potential is calculated via self-gravity of the model particles. The potential is thus altered by the changing particle masses induced by the M2M algorithm. In Paper 1 we apply \sc{primal }\rm to the target system of a smooth axisymmetric disc created by $N$-body simulations, and demonstrate that \sc{primal }\rm can reproduce the density and velocity profiles of the target system well, even when starting from a disc whose scale length is different from the target system.



We are encouraged by the success of Paper 1 and intend to build upon it. In this paper we apply \sc{primal }\rm to barred disc galaxies again generated by $N$-body simulations with GCD+ \citep{KG03,KOGBC13}. We introduce a new form of velocity observable constraints as described in \cite{DeL08}, based on the likelihood function as described in \cite{RK01}. We also introduce a rotating reference frame in a similar, although not identical fashion to \cite{LMIII}. 

We use target systems whose information is known without any error. Ultimately we wish to apply \sc{primal }\rm to real observational data, where the information will be provided for a limited region of the sky, with a more complicated selection and error function due to the dust extinction, crowding and stellar populations. However, we think that in the development stages it is important to test the algorithm against the ideal target. In this paper we demonstrate the successful application of \sc{primal }\rm to the barred galaxy targets, and this is a significant step forward to modelling the Milky Way with M2M. 

This paper is organised as follows. Section \ref{M2M} describes the M2M methodology of our original M2M algorithm as shown in Paper 1. Sections \ref{lM2M} and \ref{RF} describe the new improvements we have made to the algorithm, and Section 2.4 describes the set up of our target systems. Section \ref{R} shows the performance of our updated method for recreating the target disc systems. In Section \ref{SF} we provide a summary of this work.

\section{The M2M Algorithm: PRIMAL}
\label{PRIMAL}

The M2M algorithm works by calculating observable properties from the model and the target, and then adapting particle masses such that the properties of the model reproduce those of the target. The target can be in the form of a distribution function, an existing simulation, or real observational data. The model can be a test particle simulation in an assumed fixed or adaptive potential, or a self-gravity $N$-body model.

\subsection{A particle-by-particle M2M}
\label{M2M}
We have presented a full description of both the original M2M and our particle-by-particle M2M in Paper 1. In this section we describe briefly the basis of our particle-by-particle M2M. As mentioned in Section \ref{intro-sec}, our ultimate target is the Milky Way, and the observables are not binned data, but the position and velocity of the individual stars which are distributed rather randomly. To maximise the available constraints, we evaluate the observables at the position of each star and compare them with the $N$-body model, i.e. in a particle-by-particle fashion. To this end \sc{primal }\rm uses a kernel often used in Smoothed Particle Hydrodynamics (SPH), $W(r,h)$, which is a spherically symmetric spline function given by
\begin{equation}
\begin{array}{l}
W(r,h) = \frac{8}{\pi h^{3}} 
 \times \left\{ \begin{array}{cc}
 1-6(r/h)^{2}+6(r/h)^{3} & {\rm if}\ 0\leq r/h\leq 1/2,  \\
 2[1-(r/h)]^{3}      & {\rm if}\ 1/2\leq r/h\leq 1,  \\
 0               & {\rm otherwise},
\end{array} \right.\\
\end{array} 
\label{Weq}
\end{equation}
\citep{ML85}, where $r = \mid\textbf{r}_i-\textbf{r}_j\mid$ and $h$ is a smoothing length described later. Note that in our particle-by-particle M2M algorithm the kernel, $W(r,h)$, does not explicitly include the total mass, $M_{\text{tot}}$, unlike standard M2M algorithms, because we wish to eventually apply it to the Milky Way, whose mass is unknown. 

We use this kernel to calculate the density at the target particle locations, $\textbf{r}_j$, of both the target and the M2M model. For example, the density of the target at $\textbf{r}_j$ is evaluated by,
\begin{equation}
\rho_{t,j}=\sum_{k=1}^{N}m_{t,k}W(r_{kj},h_{j}),
\label{trho}
\end{equation}
where $m_{t,k}$ is the mass of the target particle, $r_{kj} = \mid \textbf{r}_{k} - \textbf{r}_{j} \mid$, and $h_{j}$ is the smoothing length determined by 
\begin{equation}
h_{j} = \eta \left(\frac{m_{t,j}}{\rho_{t,j}}\right)^{1/3},
\label{smoothing}
\end{equation}
where $\eta$ is a parameter and we have set $\eta=3$. In SPH simulations, a value of $\eta$ between 2 and 3 is often used. We choose the higher value to maximise the smoothness. This results in $\approx113$ particles being included in the smoothing length when the particles are distributed homogeneously in three-dimensional space. The solution of equation (\ref{smoothing}) is calculated iteratively until the relative change between two iterations is smaller than $10^{-3}$ \citep{PM07}. Similarly, the density at $h_j$ is calculated by
\begin{equation}
\rho_{j}=\sum_{i=1}^{N}m_{i}W(r_{ij},h_{j}),
\label{rho}
\end{equation}
from the model particles. The target density $\rho_{t,j}$ is calculated only once at the beginning of the M2M simulation, and the model density $\rho_{j}$ is recalculated at every timestep. We then calculate the difference between these observables
\begin{equation}
\Delta_{\rho_j} = \frac{\rho_{j}(t) - \rho_{t,j}}{\rho_{t,j}}.
\label{od2}
\end{equation}
Following the M2M algorithm, the mass of the model particle is changed to reduce $\Delta_{\rho_j}$. Thus an example of the particle mass change equation, when using density observables is:
\begin{eqnarray}
\frac{d}{dt}m_{i}(t) =  &-& \epsilon m_{i}(t)\Biggl[M\sum_{j} \frac{W(r_{ij},h_j)}{\rho_{t,j}}\tilde{\Delta}_{j,\rho}(t) \nonumber \\ 
&+& \mu \left(\ln \left(\frac{m_{i}(t)}{\hat{m}_{i}}\right)+1\right) \Biggr],
\label{WC4}
\end{eqnarray}
where $\hat{m}_i$ is the prior, and $M$ is an arbitrary constant mass, which we set as $M=10^{12}$ M$_{\sun}$ for this paper. We introduce this arbitrary constant mass so that the method may be applied to a system with unknown mass and as a result the parameters, such as $\epsilon$ and $\mu$, must be calibrated before performing the modelling. In equation (\ref{WC4}) $\tilde{\Delta}_{j,\rho}(t)$ corresponds to a temporally smoothed version of $\Delta_{\rho_j}$ to reduce fluctuations. This is derived by
\begin{equation}
\tilde{\Delta}_{j}(t) = \alpha\int_{0}^{\infty}\Delta_{j}(t-\tau)\text{e}^{-\alpha\tau}d\tau,
\end{equation}
with $\alpha$ being a small positive parameter. This $\tilde{\Delta}_{j}(t)$ can be calculated from the differential equation
\begin{equation}
\frac{d\tilde{\Delta}(t)}{dt} = \alpha(\Delta-\tilde{\Delta}).
\label{NewDel2}
\end{equation}
This temporal smoothing effectively increases the number of particles from $N$ to
\begin{equation}
N_{\text{eff}}=N\frac{t_{1/2}}{\Delta t},
\end{equation}
where $\Delta t$ is the length of the timestep and $t_{1/2}=(\text{ln }2)/\alpha$ is the half life of the ghost particles. \cite{ST96} show that excessive temporal smoothing is undesirable, and should be limited to $\alpha\geq2\epsilon$. The regularisation term, $\mu \left(\ln \left(\frac{m_{i}(t)}{\hat{m}_{i}}\right)+1\right)$, in equation (\ref{WC4}) is introduced to stabilise the mass changing process. The idea behind this term is to maximise the entropy defined by
\begin{equation}
S = - \sum_{i} m_{i} \ln\left(\frac{m_{i}}{\hat{m}_{i}}\right),
\label{S}
\end{equation}
and $\mu$ is a parameter to control the regularization.

In this paper, the prior, $\hat{m}_i$, is set as $\hat{m}_{i}=M_{\text{tot,ini}}/N$, where $M_{\text{tot,ini}}$ is the initial total mass of the system and $N$ is the number of particles. The regularization term forces the particle mass close to their initial value. As with Paper 1, we write $\epsilon$ as $\epsilon = \epsilon'\epsilon''$ where $\epsilon''$ is given by 
\begin{equation}
\epsilon'' = \frac{10}{\text{max}_{i}\left(M\sum_{j}\frac{W(r_{ij},h_j)}{\rho_{j,t}}\tilde{\Delta}_{\rho_j}(t)\right)}.
\label{e''}
\end{equation}

\subsection{Maximum likelihood for velocity constraints} 
\label{lM2M}
In Paper 1 we use velocity observables in the form of mean local velocity field, calculated around the target particle positions, with the kernel described in equation (\ref{Weq}). However as the Galactic stellar surveys will provide us velocity information for individual particles, instead of smoothing the velocity, we can evaluate likelihood of the actual velocity of the particle. Thus we have converted the velocity section of our algorithm to maximise the likelihood of the velocity of the target particles as shown in \cite{DeL08}. The likelihood is calculated with the equation
\begin{equation}
\mathcal{L}=\sum_{j}\text{ln}(\mathcal{L}_{j}),
\label{L}
\end{equation}
where $\mathcal{L}_j$ is the likelihood function for a single discrete velocity. Following \cite{RK01}, we calculate the likelihood for individual velocity observables, $v_j$, at the target particle positions, $\textbf{r}_j$, with 
\begin{equation}
\mathcal{L}_{j}(v_{j},\textbf{r}_{j})=\frac{1}{\sqrt{2\pi}}\int\left(\frac{\text{d}L}{\text{d}v}\right)_j\text{e}^{-(v_{j}-v)^2/2\sigma^2_j}\text{d}v,
\label{L2}
\end{equation}
where $\sigma_{j}$ is the velocity error, which we have set as $\sigma_j=2.5\text{ km s}^{-1}$ for this paper, and $\text{d}L/\text{d}v$ is a velocity distribution for the model. Although we fix the velocity error, and do not discuss the effects of the errors in this paper, an advantage of the likelihood-based velocity constraints is that we can set individual errors for each velocity component of each particle. Instead of the kernel chosen in \cite{DeL08} we use our kernel from equation (\ref{Weq}), allowing us to write $\text{d}L/\text{d}v$ for target particle $j$, from model particle $i$, as
\begin{equation}
\left(\frac{\text{d}L}{\text{d}v}\right)_{j}=\frac{1}{l_{j}}\sum_{j}W_{ij}m_{i}\delta(v_{j}-v_{i}),
\end{equation}
where $\delta(x)$ is the delta function, and 
\begin{equation}
l_{j}=\sum_{i}W_{ij}m_{i},
\end{equation}
which is the same as equation (\ref{rho}). We can express $\mathcal{L}_{j}$ in equation (\ref{L2}) as
\begin{equation}
\mathcal{L}_{j}=\frac{\hat{\mathcal{L}}_{j}}{l_{j}},
\end{equation}
where 
\begin{equation}
\hat{\mathcal{L}}_{j}=\frac{1}{\sqrt{2\pi}}\sum_{i}W_{ij}m_{i}\text{e}^{-(v_{j}-v_{i})^2/2\sigma^2_j},
\end{equation}
and
\begin{equation}
\frac{\text{d}\hat{\mathcal{L}_j}}{\text{d}m_{i}}=\frac{1}{\sqrt{2\pi}}W_{ij}\text{e}^{-(v_{j}-v_{i})^2/2\sigma_j^2}.
\end{equation}
This leads us to the modified term in the particle mass change equation. Following the M2M algorithm, we maximise the likelihood of equation (\ref{L}),  using
\begin{equation}
\frac{\text{d}m_{i}}{\text{d}t}=\epsilon m_{i}M\frac{\text{d}\mathcal{L}_j}{\text{d}m_{i}},
\end{equation}
where
\begin{eqnarray}
\frac{\text{d}\mathcal{L}_j}{\text{d}m_{i}}&=&\frac{\text{d}}{\text{d}m_{i}}\sum_j\text{ln}\left(\frac{\hat{\mathcal{L}}_{j}}{l_{j}}\right) \nonumber \\ 
&=& \sum_j\left[\frac{\text{d}}{\text{d}m_{i}}\text{ln}(\hat{\mathcal{L}}_{j})-\frac{\text{d}}{\text{d}m_{i}}\text{ln}(l_{j})\right] \nonumber \\
&=& \sum_j\left[\frac{1}{\hat{\mathcal{L}}_{j}}\frac{\text{d}}{\text{d}m_{i}}\hat{\mathcal{L}}_{j}-\frac{1}{l_{j}}W_{ij}\right].
\end{eqnarray}
The particle mass change equation from the velocity based likelihood constraints is calculated with
\begin{equation}
\frac{\text{d}m_{i}}{\text{d}t}=\epsilon m_{i}M\sum_{j}W_{ij}\left[\frac{1}{\sqrt{2\pi}}\frac{\text{e}^{-(v_{j}-v_{i})^2/2\sigma^2_j}}{\hat{\mathcal{L}_j}}-\frac{1}{l_{j}}\right],
\end{equation}
and equation (\ref{WC4}) becomes 
\begin{eqnarray}
\frac{d}{dt}m_{i}(t) &=& -\epsilon m_{i}(t)\Biggl\{M\sum_{j} \frac{W(r_{ij},h_j)}{\rho_{t,j}}\Delta_{j}(t) \nonumber \\  
&-& \zeta M\Biggl[\sum_{j}W_{ij}\left(\frac{1}{\sqrt{2\pi}}\frac{\text{e}^{-(v_{r,j}-v_{r,i})^2/2\sigma^2_{r,j}}}{\hat{\mathcal{L}_{r,j}}}-\frac{1}{l_{j}}\right) \nonumber \\
&+& \sum_{j}W_{ij}\left(\frac{1}{\sqrt{2\pi}}\frac{\text{e}^{-(v_{z,j}-v_{z,i})^2/2\sigma^2_{z,j}}}{\hat{\mathcal{L}_{z,j}}}-\frac{1}{l_{j}}\right) \nonumber \\
&+& \sum_{j}W_{ij}\left(\frac{1}{\sqrt{2\pi}}\frac{\text{e}^{-(v_{\text{rot},j}-v_{\text{rot},i})^2/2\sigma^2_{\text{rot},j}}}{\hat{\mathcal{L}_{\text{rot},j}}}-\frac{1}{l_{j}}\right)\Biggr] \nonumber \\
&+&  \mu \left(\ln \left(\frac{m_{i}(t)}{\hat{m}_{i}}\right)+1\right)\Biggr\},
\end{eqnarray}
where $v_r$, $v_z$ and $v_{\text{rot}}$ are the radial, vertical and rotational velocity components. The parameter, $\zeta$, is an optional adjustable parameter for changing the significance of the velocity constraints, although we set $\zeta=1$ in this paper. Following \cite{DeL08}, we use temporally smoothed versions (c.f. equation \ref{NewDel2}) of $\hat{\mathcal{L}}$ and $l$.

Alongside this new mass change equation we have also altered the time when the constraints kick in from our previous method. We found when using the likelihood-based velocity constraints the model requires a lower level of temporal smoothing, and thus we are able to use temporal smoothing as soon as the mass change equation is enabled. Thus we now use the following series of stages: From $t=$ 0 to 0.471 Gyr (one simulation time unit) we allow the initial model to experience relaxation, following a standard self-gravity $N$-body calculation without any mass change. From $t=$ 0.471 Gyr to 0.942 Gyr we used temporally smoothed density constraints only, and at $t=0.942$ Gyr we engage the velocity constraints as well. This sequence is substantially shorter than the method used in Paper 1, allowing the solution to converge faster, and the overall simulation length to be halved to $\sim5$ Gyr. We continue to use individual timesteps for the particles, and only update the masses of particles whose position and velocity are updated within the individual timestep. The timestep for each particle is determined by
\begin{equation}
dt_i=C_{\text{dyn}}\left(\frac{0.5h_i}{\vert d\textbf{v}_i/dt\vert}\right)^{\frac{1}{2}},
\end{equation}
with $C_{\text{dyn}}=0.2$. We also retain the limit on the maximum mass change which any particle can experience in one timestep. We set this limit to ten percent of that particles mass.

We have again performed a parameter search to determine differences in the likelihood-based velocity constraints, as we did in Paper 1. There are four important parameters, $\epsilon$, $\alpha$, $\zeta$ and $\mu$, which must be calibrated for M2M. $\epsilon$ provides the balance between the speed of convergence, and the smoothness of the process. Note that $\epsilon=\epsilon'\epsilon''$ where $\epsilon''$ is defined by equation (\ref{e''}). We have chosen $\epsilon'=0.1$ as an appropriate balance between accuracy and simulation time. With more computing power available to us we would consider running a lower value of $\epsilon$. However if $\epsilon'\ll0.1$, it is possible the model might not converge as the mass change is too slow. The choice of $\alpha$, which controls the strength of the temporal smoothing, should depend upon the choice of $\epsilon$ ($\alpha\geq2\epsilon$). We find that our modelling is not overly sensitive to $\alpha$ as long as the condition $\alpha\geq2\epsilon$ is met and we set $\alpha=2.0$ in this paper. We set $\zeta=1$ as mentioned before. $\mu$ controls the strength of the regularization and is essential in reducing the oscillation in particle masses and ensuring smooth convergence. We discuss the importance of $\mu$ in much greater detail in Paper 1. In this paper we adopt $\mu=10^5$. All different models presented in Section \ref{R} use this same parameter set, and have not been individually tailored to the target or model in question. This demonstrates the robustness of the method.

\subsection{Rotating reference frames}
\label{RF}
In Paper 1 it was sufficient to use a fixed reference frame as we were investigating smooth axisymmetric discs. However, if the target has some non-axisymmetric structure, such as a bar, the target bar angle is fixed, but the bar of the model rotates in the fixed reference frame. For example, if there is a bar, we expect the density and kinematics to be very different at the different azimuth angles at a fixed radius. Then, if the bar of the model is not aligned with the target bar, the observables of the model are evaluated in the different dynamical states from the target observables. Hence, if the bar of the model keeps rotating in the fixed frame, the model particles receive the different constraints from the target depending on the bar angle at each timestep, and the model never settles to the solution. This is discussed in Section \ref{R}.

\cite{LMIII} have proposed using a reference frame with a fixed bar angle, and comparing multiple simulations with different bar angles to find the best fit. This was trivial for their model, because they used a fixed shape of the bar potential and rotating with a fixed bar pattern speed, $\Omega_p$. We however have not assumed any pattern speed prior to the beginning of our simulations, nor have we placed any explicit constraints on it. Instead we start with a smooth disc as the initial condition, allowing the pattern speed to evolve with the model galaxy due to self-gravity, and compare $\Omega_p$ for the model and target galaxies at the end of the run.

We therefore calculate the angle of the bar in our target, and the angle of the bar in the model at each step. Then we rotate the model to match the bar angle of the target for the purposes of calculating the observables in the same reference frame. It is our hope that this method will allow the pattern speed to be recovered along with the density and velocity profiles. When applying this to the Milky Way we will not know the exact bar angle. But here, we assume that the bar angle is known for our first step of modelling the bar. We call this reference frame change the $rotating$ $reference$ $frame$ hereafter. In Section \ref{R} we present a comparison of our method with and without this rotating reference frame, and also present the results from cases where we have chosen an incorrect bar angle.

\begin{figure}
\centering
\resizebox{\hsize}{!}{\includegraphics{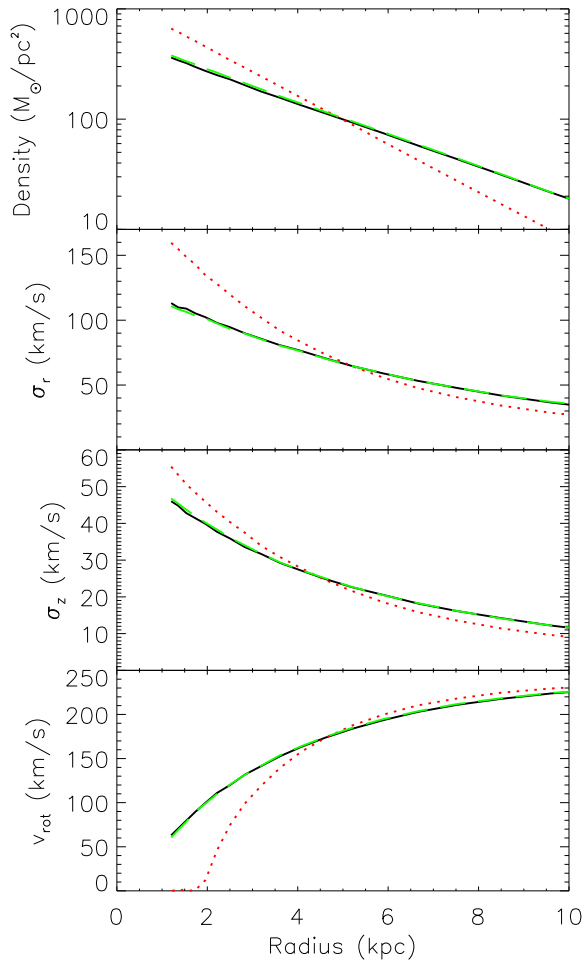}}
\caption{Initial (red dotted), final (green dashed) and target (black solid) density profile (upper), radial velocity dispersion (upper middle), vertical velocity dispersion (lower middle) and rotation velocity (lower) for Model A with Target I.}
\label{AG}
\end{figure}

\begin{figure}
\centering
\resizebox{\hsize}{!}{\includegraphics{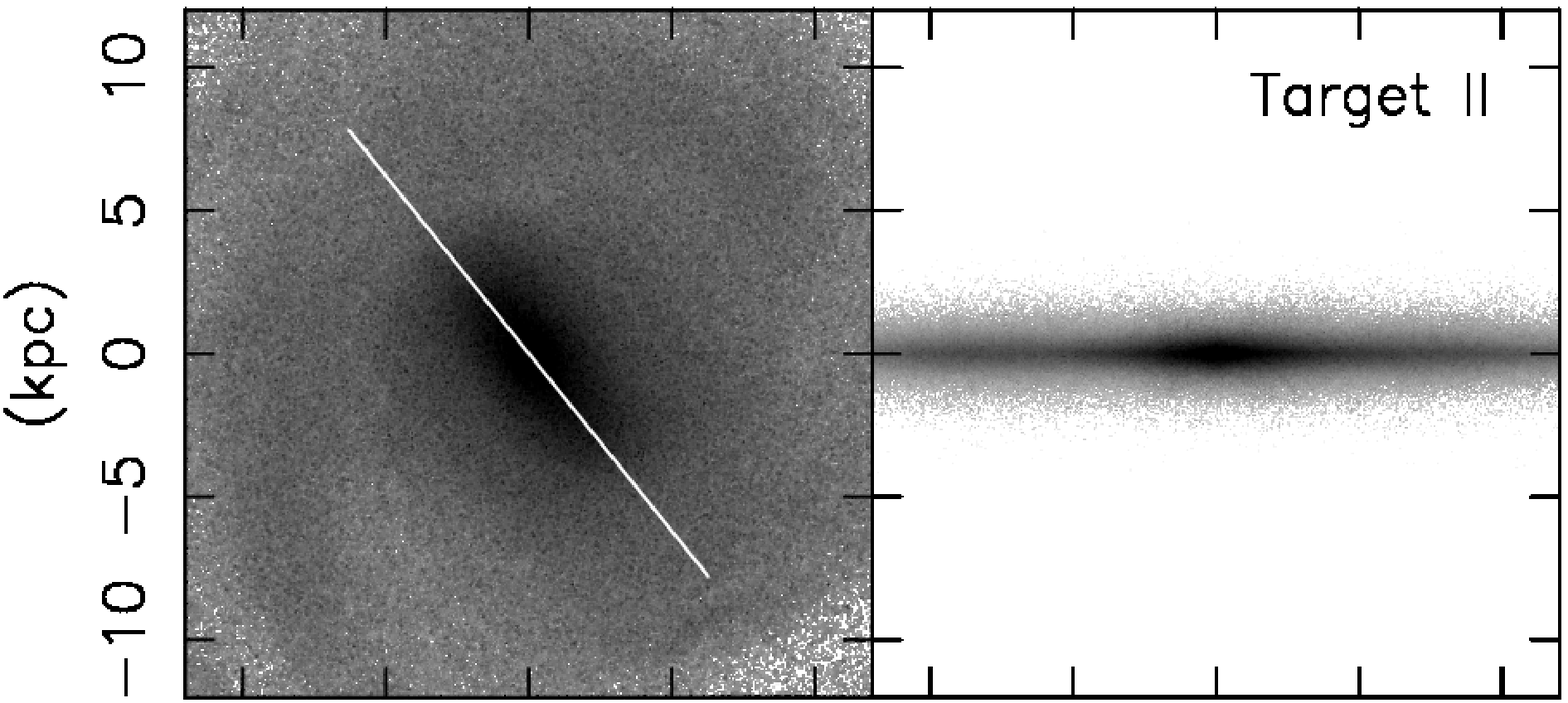}}
\resizebox{\hsize}{!}{\includegraphics{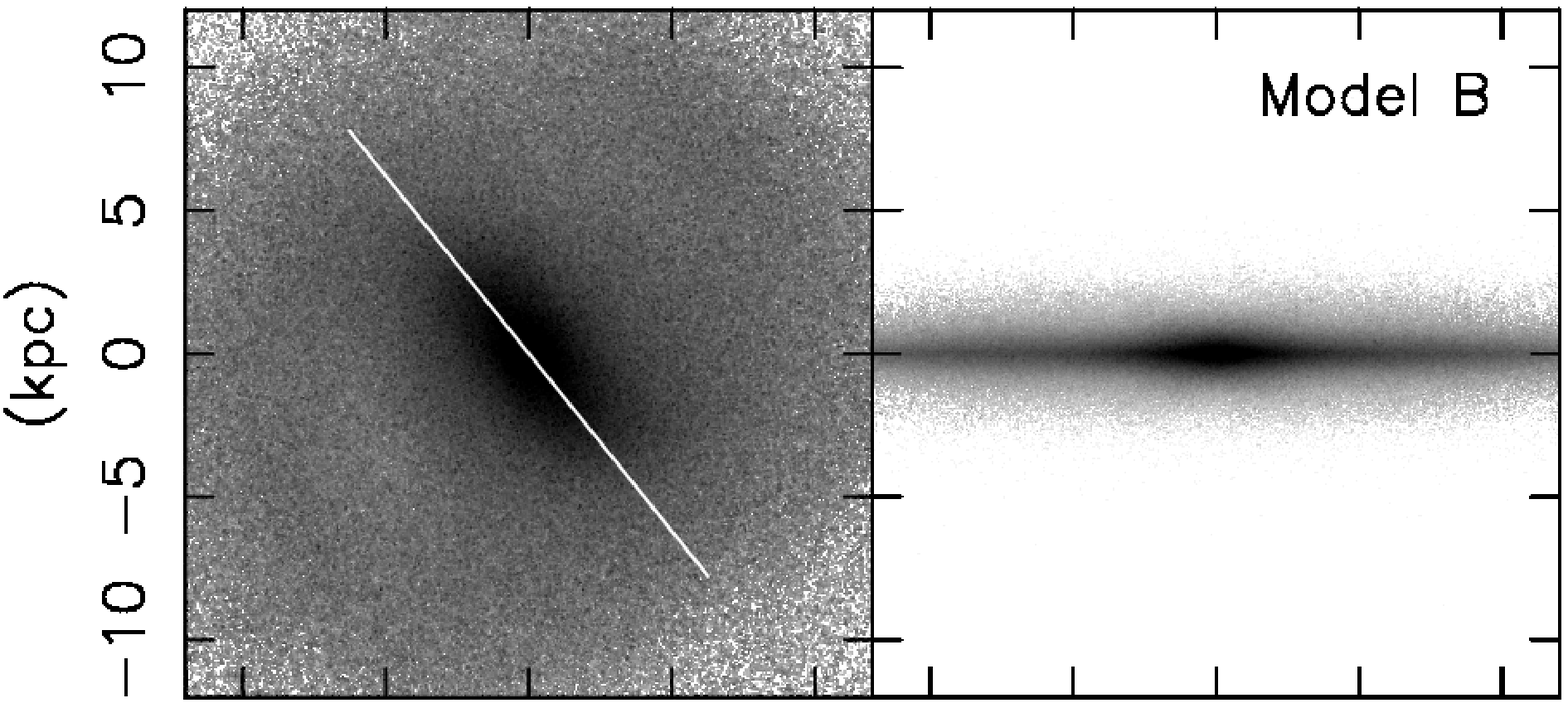}}
\resizebox{\hsize}{!}{\includegraphics{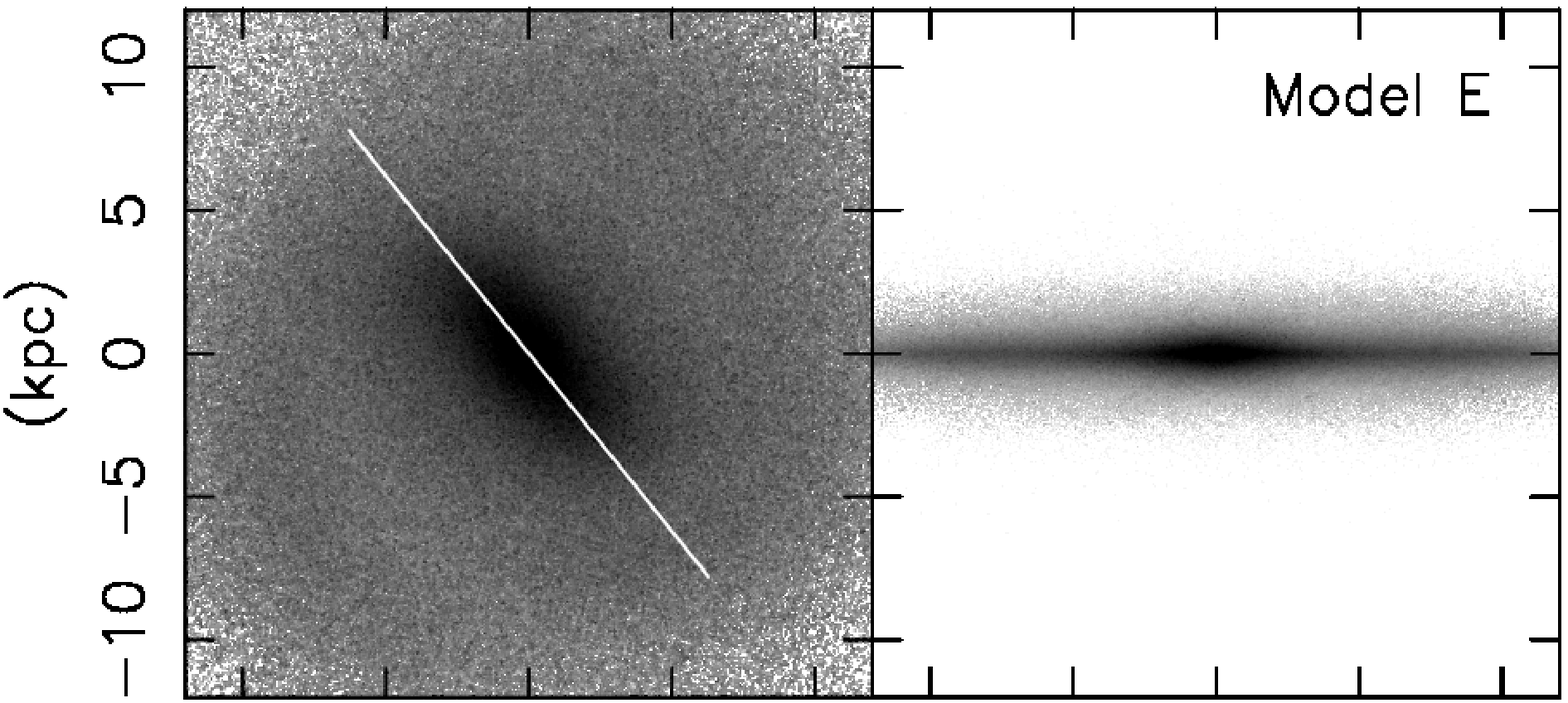}}
\resizebox{\hsize}{!}{\includegraphics{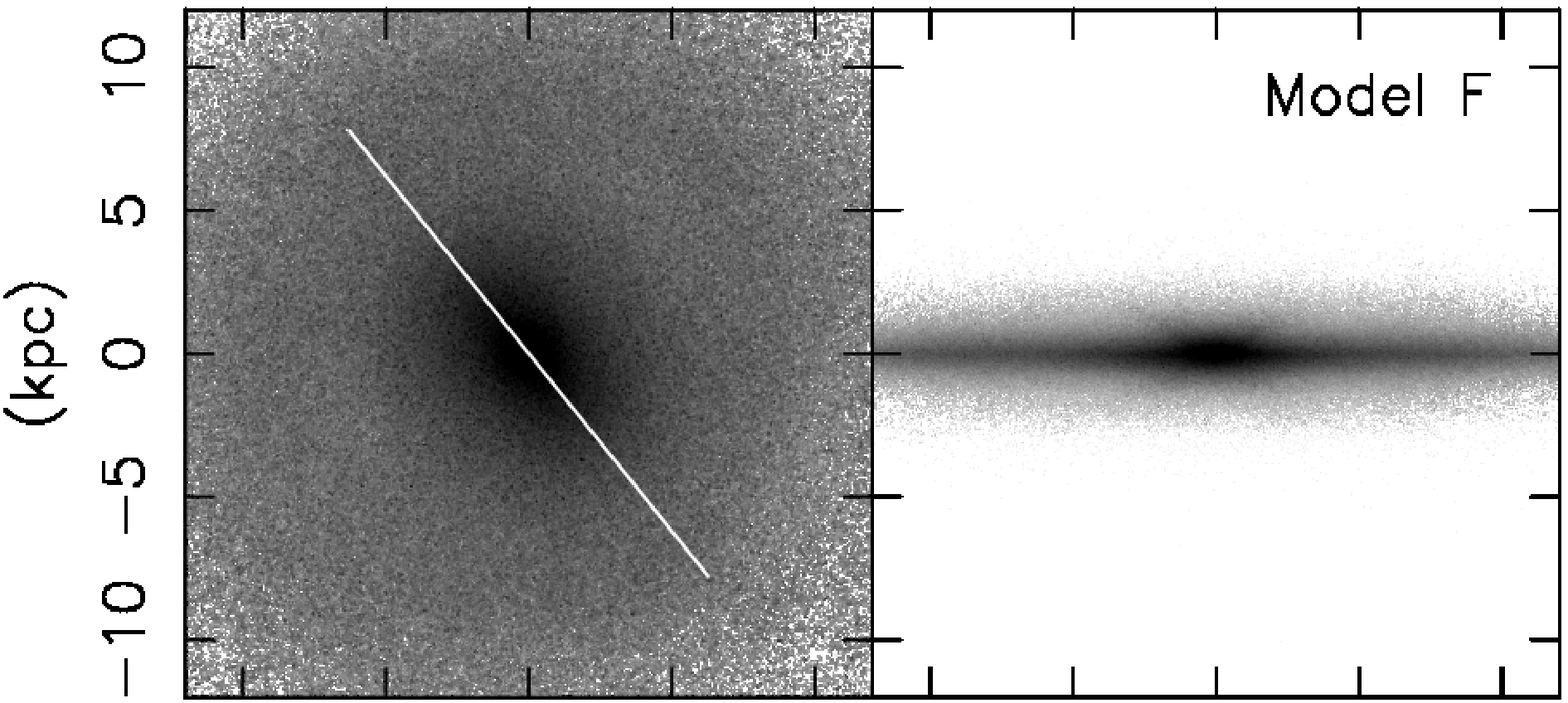}}
\resizebox{\hsize}{!}{\includegraphics{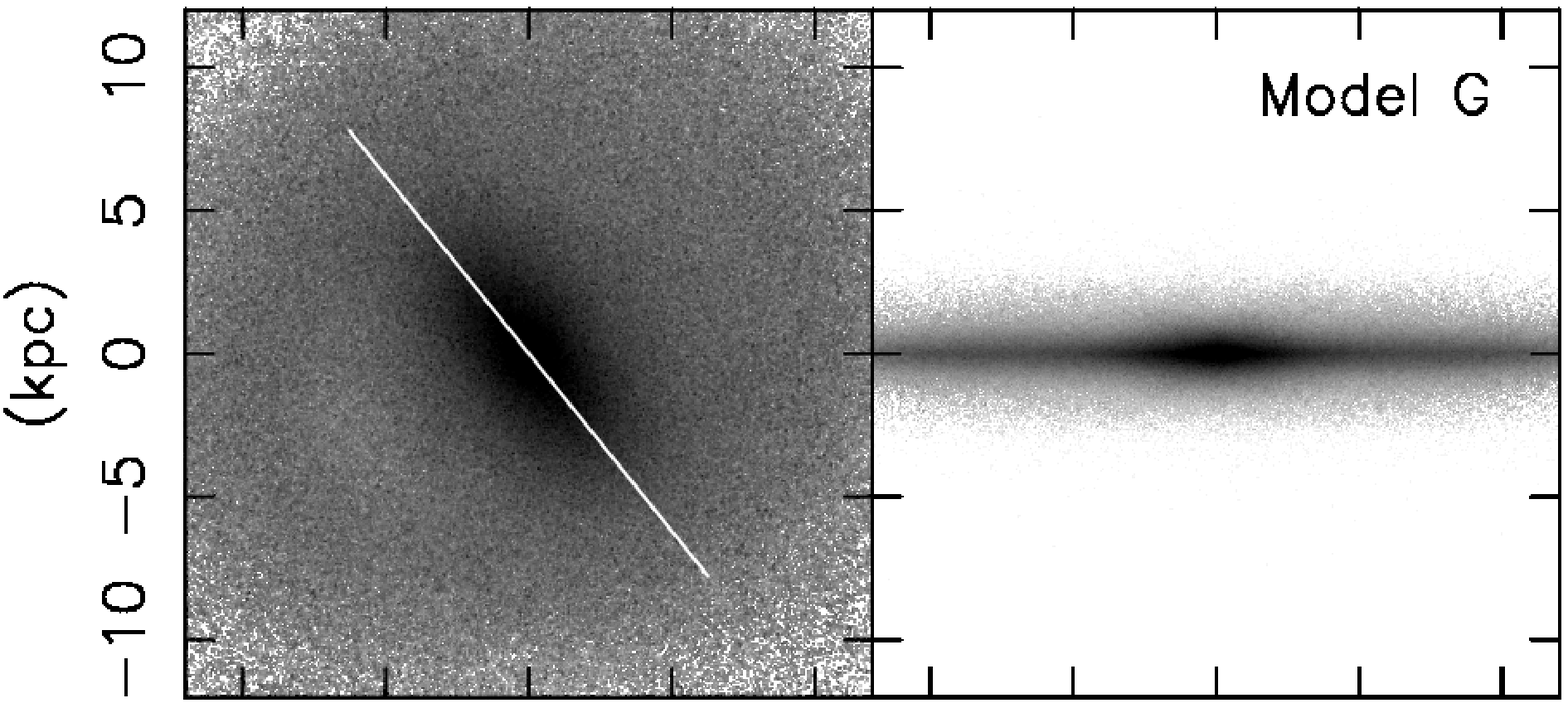}}
\resizebox{\hsize}{!}{\includegraphics{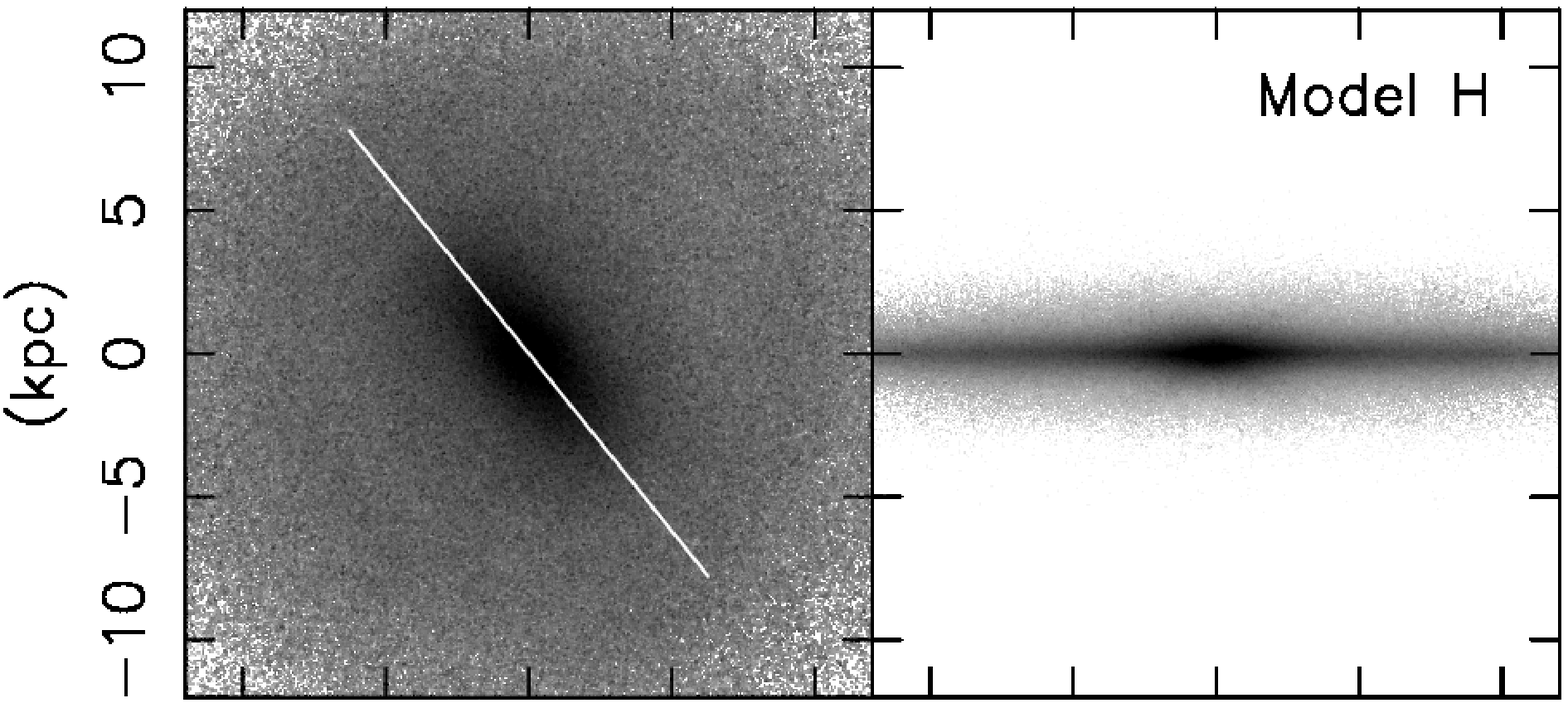}}
\caption{face-on (left) and edge-on (right) density maps of Target II, Models B, E, F, G and H, (from top to bottom) plotted for comparison. The white line indicates the angle of the bar, rotated for comparison. The density scale is the same for all panels.}
\label{AO}
\end{figure}
\begin{figure}
\centering
\resizebox{\hsize}{!}{\includegraphics{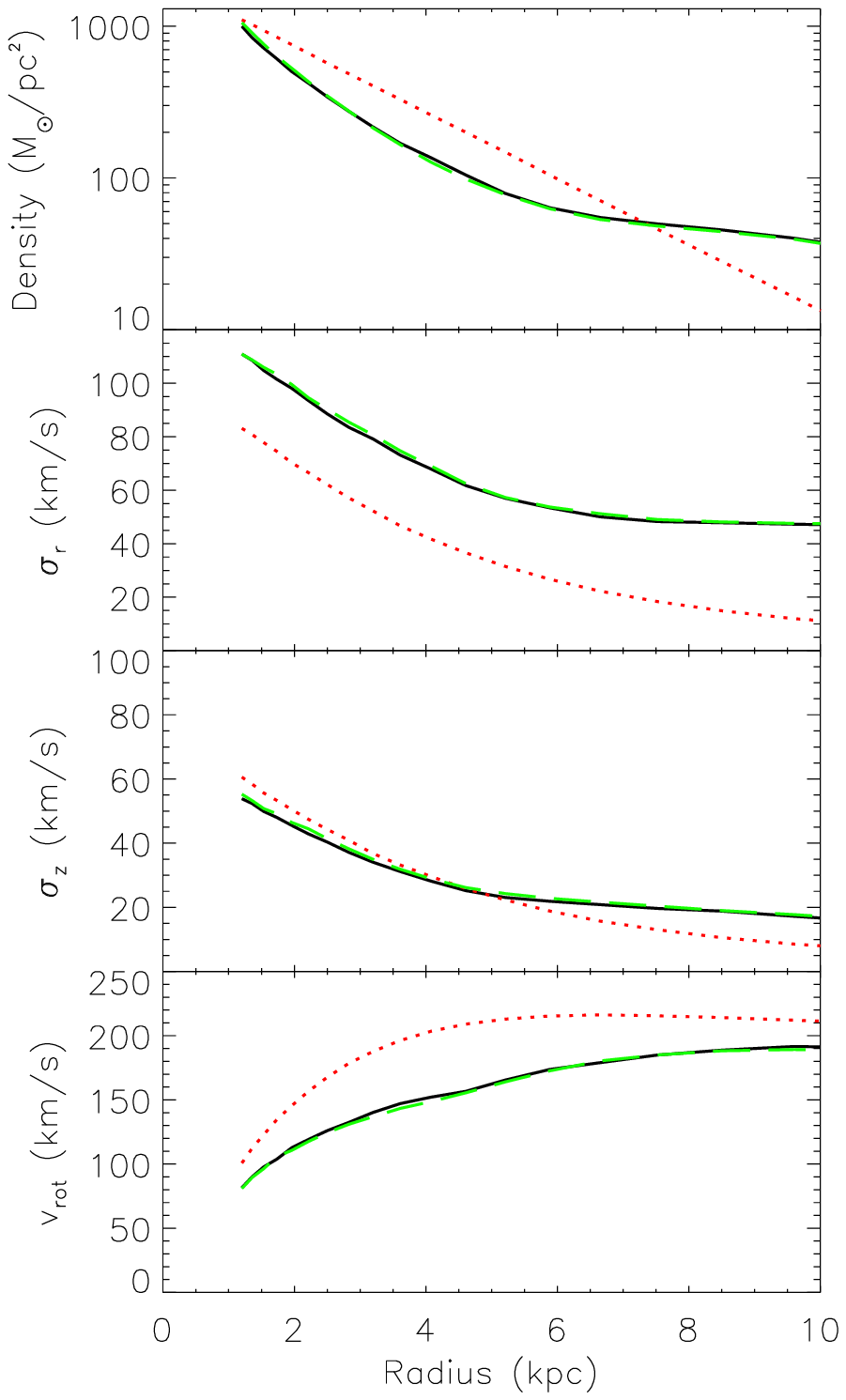}}
\caption{Same as Fig. \ref{AG}, but for Model B with Target II.}
\label{BG}
\end{figure}
\subsection{Target system setup}
\label{p2pSetup}
Our simulated target galaxies consist of a pure stellar disc with bar and/or spiral structure and a static dark matter halo, set up using the method described in \cite{GKC12}. The dark matter halo density profile is taken from a truncated NFW profile \citep{NFW97,RAS08} and given by;
\begin{table*}
\centering
\caption{$N$-body target parameters. $M_{200}$ is the mass of the halo, $M_d$ is the mass of the disc, $c$ is the concentration parameter, $z_d$ is the scale height,  $\sigma_r^2/\sigma_z^2$ is the ratio between radial and tangential velocity dispersion and $\Omega_{t,p}$ is the pattern speed of the bar, measured at 2 Gyr.}
\label{Tpars}
\renewcommand{\footnoterule}{}
\begin{tabular}{@{}cccccccccccc@{}}
\hline
Target &  $M_{200}$ ($M_{\sun}$) & $M_d$ ($M_{\sun}$) & $c$ & $z_d$ (kpc) & $\sigma_r^2/\sigma_z^2$ & $\Omega_{t,p}$ ($\text{km s}^{-1}\text{kpc}^{-1}$) & Notes \\ \hline
I & $1.75\times10^{12}$ & 3.0$\times10^{10}$ & 20.0  & 0.35 & 9.0 & N/A & Smooth Disc\\ \hline
II & $2.0\times10^{12}$ & 5.0$\times10^{10}$ & 9.0 & 0.3 & 2.0 & 27.5 \\ \hline
III & $1.5\times10^{12}$ & 5.0$\times10^{10}$ & 7.0  & 0.3 & 2.0 & 31.7 \\ \hline
IV & $1.75\times10^{12}$ & 5.0$\times10^{10}$ & 9.0  & 0.3 & 2.0 & 28.9 \\ \hline
\end{tabular}
\end{table*}
\begin{equation}
\rho_{dm}=\frac{3H_0^2}{8\pi G}\frac{\delta_c}{cx(1+cx)^2}\text{e}^{-x^2},
\label{rhodm}
\end{equation}
where $\delta_c$ is the characteristic density described by \cite{NFW97}. The truncation term, $\text{e}^{-x^2}$, is introduced in our initial condition generator for a live halo simulation. Although we use a static dark matter halo in this paper, we used the profile of equation (\ref{rhodm}). Note that the truncation term leads to very little change in the dark matter density profile in the inner region focused on in this paper. The concentration parameter $c=r_{200}/r_s$ and $x=r/r_{200}$, where $r_{200}$ is the radius inside which the mean density of the dark matter sphere is equal to $200\rho_{\text{crit}}$ and given by;
\begin{equation}
r_{200}=1.63\times10^{-2}\left(\frac{M_{200}}{h_{100}^{-1}M_{\sun}}\right)^{\frac{1}{3}}h_{100}^{-1}\text{kpc},
\end{equation}
where $h_{100}=H_0/(100 \text{ km s}^{-1}$ Mpc$^{-1})$, and $H_0$ is the Hubble constant set to $71$ km s$^{-1}$ Mpc$^{-1}$.

The stellar disc is assumed to follow an exponential surface density profile:
\begin{equation}
\rho_d=\frac{M_d}{4\pi z_dR_d^2}\text{sech}^2\left(\frac{z}{z_d}\right)\text{e}^{-R/R_d},
\end{equation}
where $z_d$ is the scale height of the disc and $R_d$ is the scale length. The velocity dispersion for each three dimensional position is computed following \cite{SMH05} to construct a near-equilibrium condition for each the target discs. We have constructed four different target galaxies whose initial conditions are listed in Table 1, and the scale length of the target discs are initially set as $R_{t,d}=3$ kpc. We run $N$-body simulations with these initial conditions, with $10^6$ particles, for 2 Gyr using a tree $N$-body code, GCD+ \citep{KG03,KOGBC13}, and adopt the final output as a target. We use the kernel softening suggested by \cite{PM07}. Although these authors suggested adaptive softening length, we use a fixed softening for these simulations for simplicity. Our softening length $\varepsilon=0.577\text{ kpc}$ is about three times larger than the equivalent Plummer softening length. We also use this softening for the M2M modelling runs. 
\begin{figure}
\centering
\resizebox{\hsize}{!}{\includegraphics{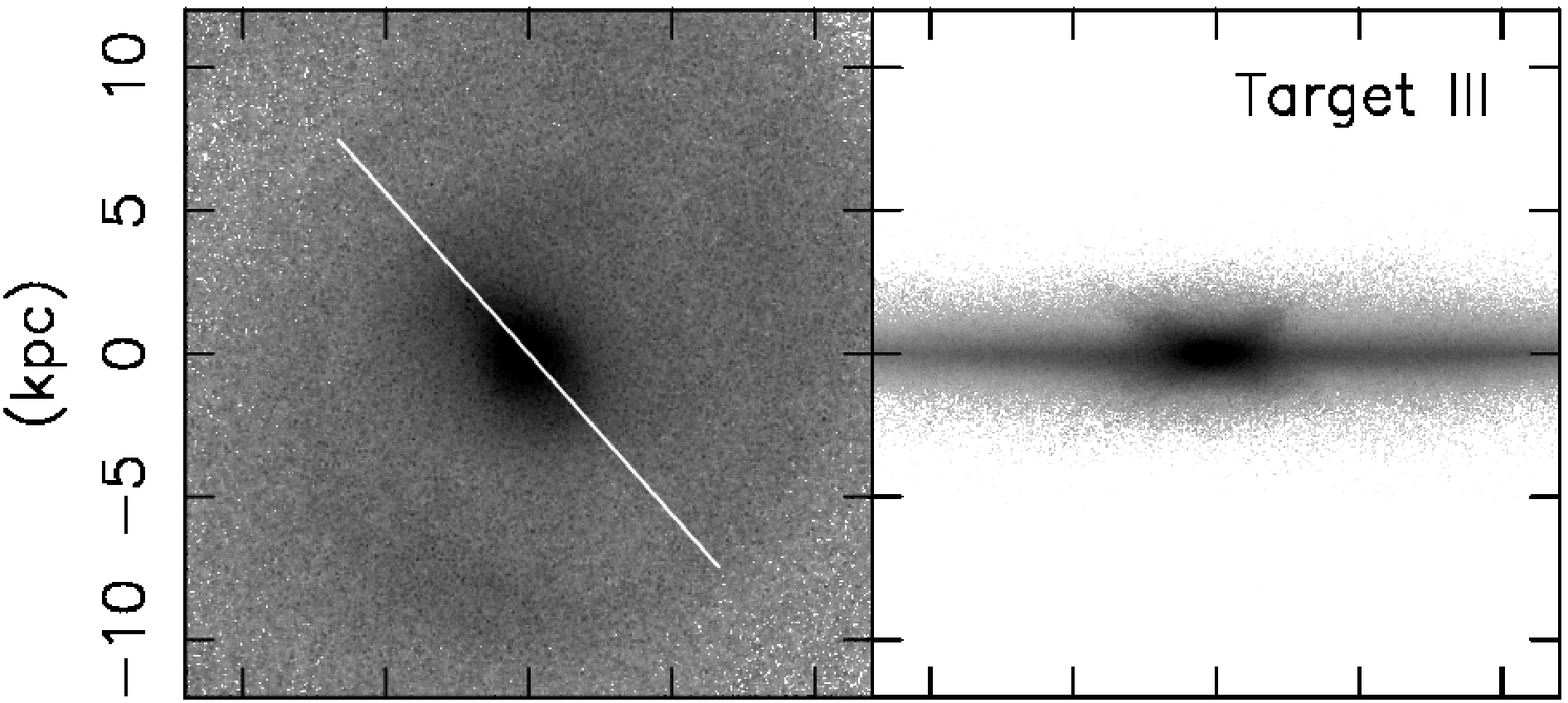}}
\resizebox{\hsize}{!}{\includegraphics{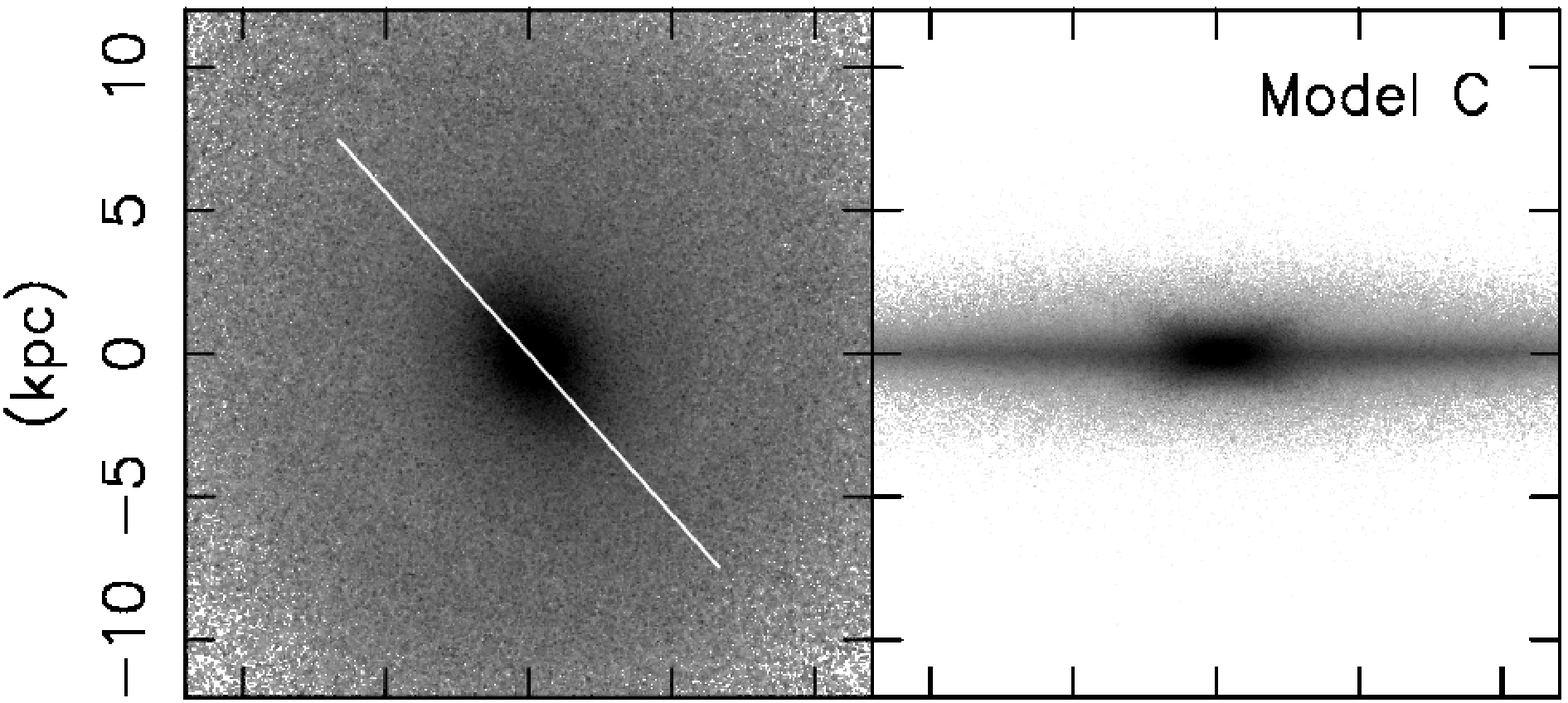}}
\caption{Same as Fig. \ref{AO}, but for Model C with Target III.}
\label{CTM}
\end{figure}
\begin{figure}
\centering
\resizebox{\hsize}{!}{\includegraphics{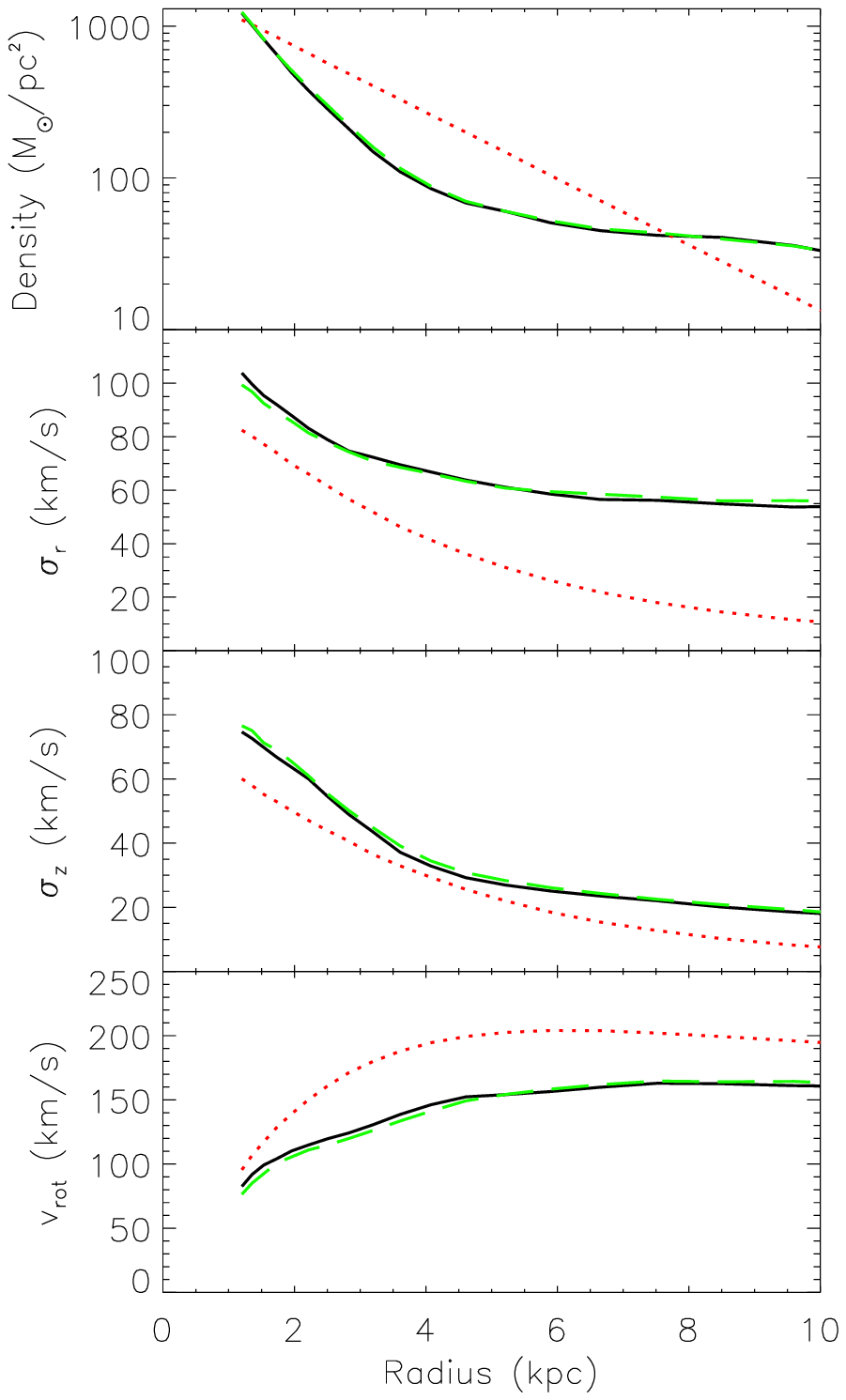}}
\caption{Same as Fig. \ref{AG}, but for Model C with Target III.}
\label{CG}
\end{figure}

\begin{table*}
\centering
\caption{M2M model results at the final timestep. $\Omega_{t,p}$ is the target pattern speed, $\Omega_p$ is the model pattern speed, $\chi^2_{\rho}$ is a measure of accuracy of the density, $\mathcal{L}_{r,z,rot}$ are the likelihood values for the radial, tangential and rotational velocity.}
\label{Mpars}
\renewcommand{\footnoterule}{}
\begin{tabular}{@{}cccccccccccc@{}}
\hline
Model & Target & $\Omega_p$ ($\text{km s}^{-1}\text{kpc}^{-1}$) & $\chi^2_{\rho}$ & $-\mathcal{L}_r/10^6$ & $-\mathcal{L}_z/10^6$ & $-\mathcal{L}_{\text{rot}}/10^6$ & Notes \\ \hline
A & I & N/A & 0.123 & 6.06 & 3.61 & 5.27 & Smooth Disc\\ \hline
B & II & 27.9 & 0.254 & 4.72 & 3.62 & 4.43  \\ \hline
C & III & 30.4 & 0.235 & 4.96 & 4.40 & 5.28 &  \\ \hline
D & IV & 28.3 & 0.189 & 5.14 & 5.15 & 5.02 & \\ \hline
E & II & 27.3 & 0.230 & 4.77 & 3.65 & 4.47 & Partial Data \\ \hline
F & II & 23.6 & 0.276 & 5.77 & 3.91 & 5.21 & Bar angle $-30^o$ \\ \hline
G & II & 28.0 & 0.250 & 4.87 & 3.67 & 4.55 & Bar angle $-10^o$ \\ \hline
H & II & 24.3 & 0.21-0.49 & 6.82-9.99 & 4.59-5.48 & 6.63-8.77 & No rotating frame \\ \hline
\end{tabular}
\end{table*}

As mentioned above, in this initial stage of development, we assume that the dark matter halo potential is known and there is no other external potential such as the bulge or stellar halo. We use the same number of particles, $10^6$, and the same initial dark matter halo and disc parameters for the model and target galaxies, except for the disc scale length: $R_d=2$ kpc for the models and $R_{t,d}=3$ kpc for the targets.
\begin{figure}
\centering
\resizebox{\hsize}{!}{\includegraphics{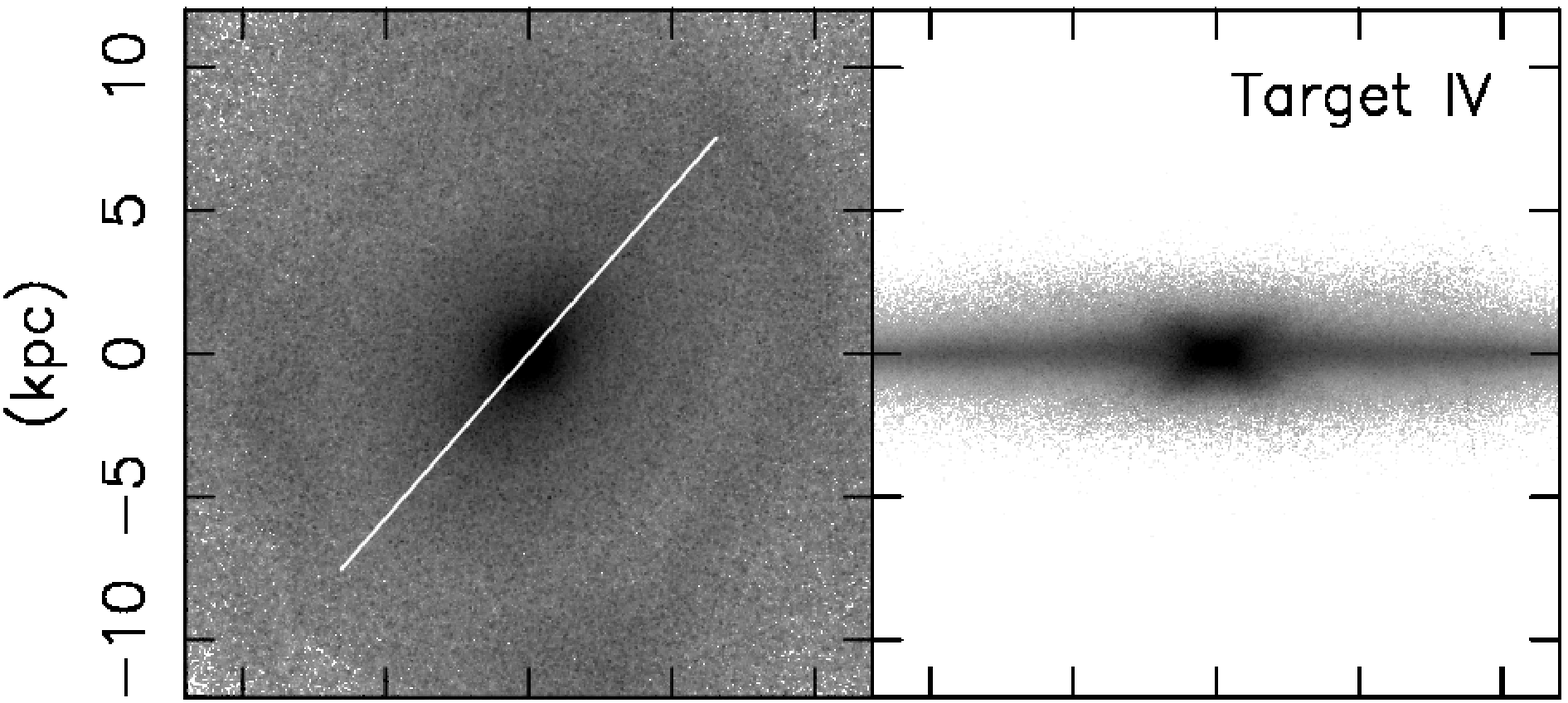}}
\resizebox{\hsize}{!}{\includegraphics{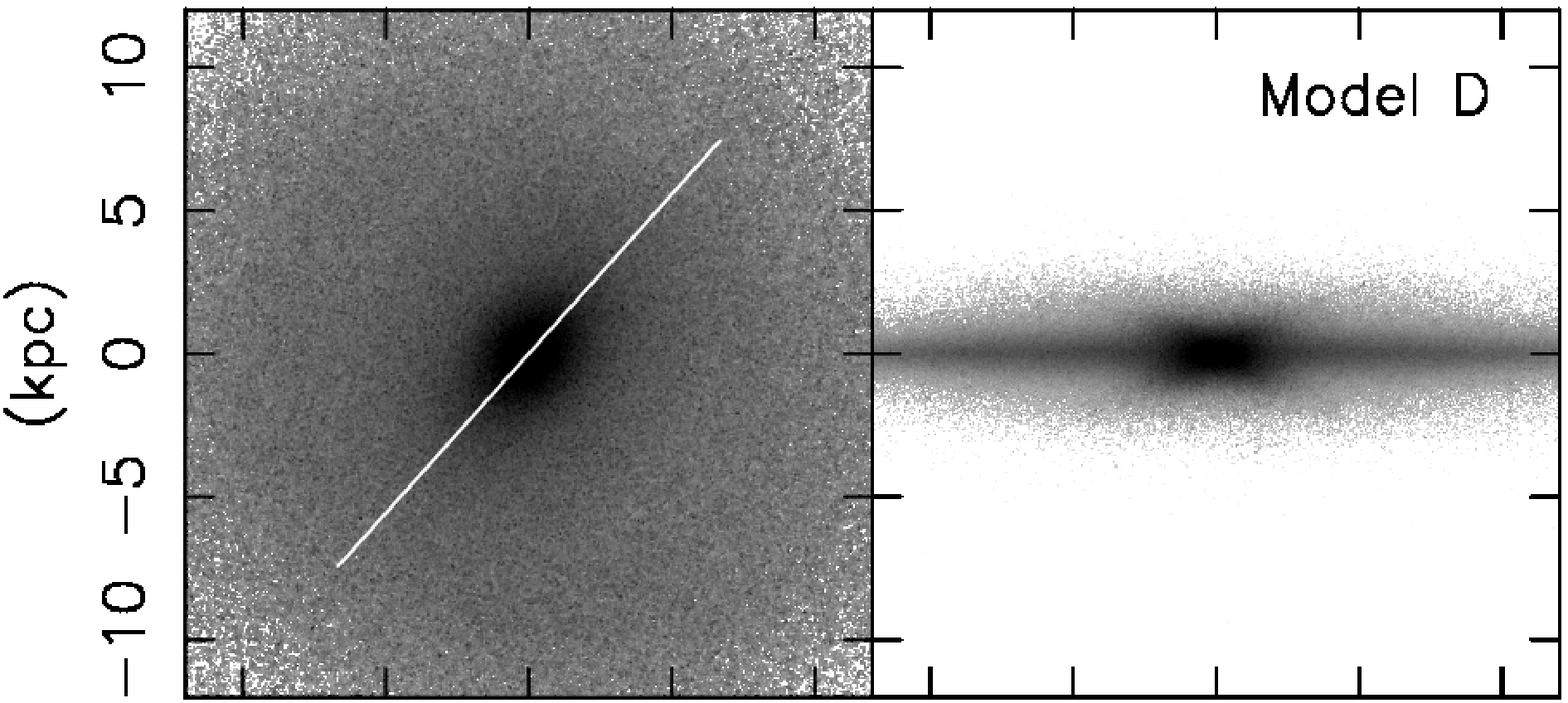}}
\caption{Same as Fig. \ref{AO}, but for Model D with Target IV.}
\label{DTM}
\end{figure}
\begin{figure}
\centering
\resizebox{\hsize}{!}{\includegraphics{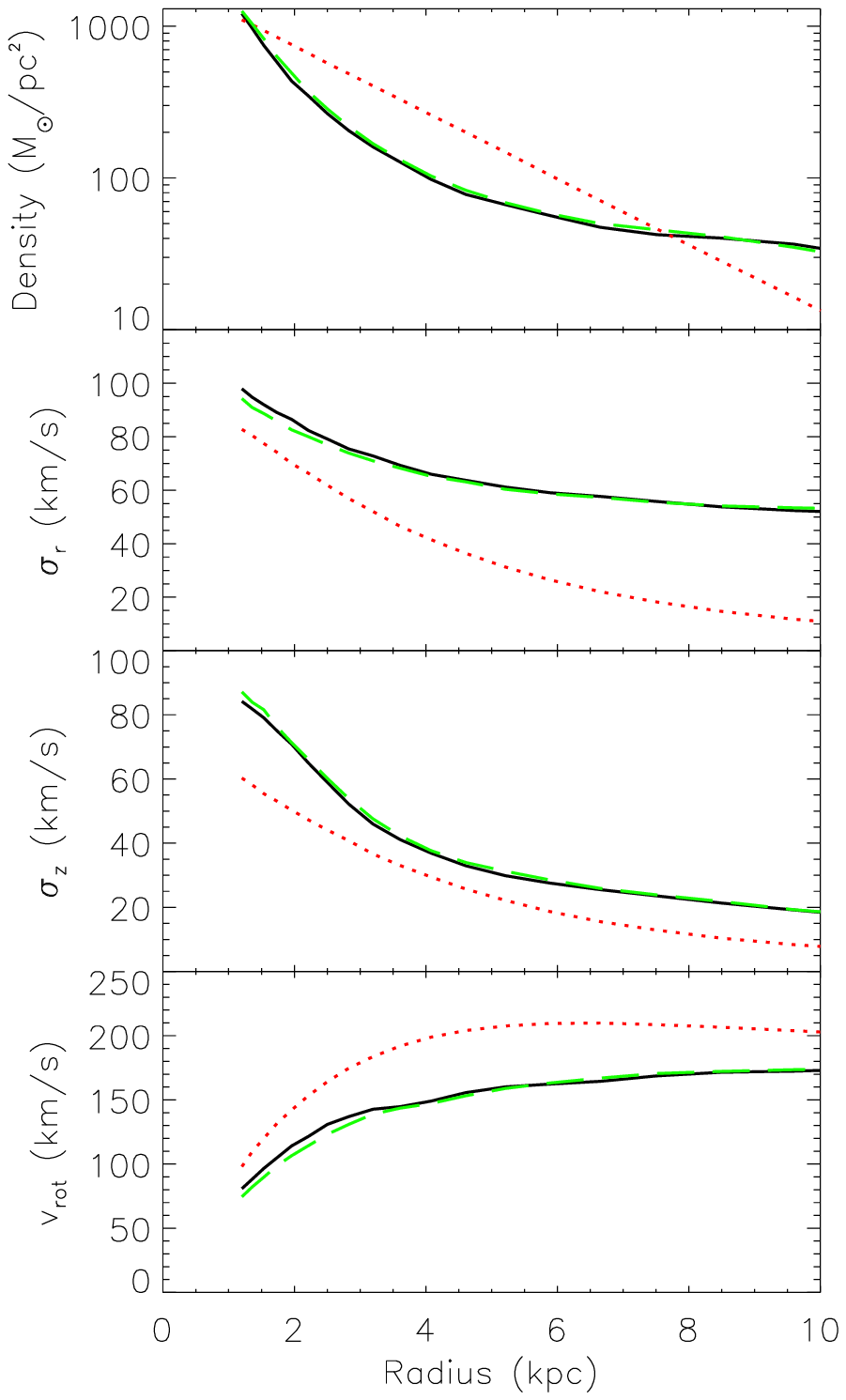}}
\caption{Same as Fig. \ref{AG}, but for Model D with Target IV.}
\label{DG}
\end{figure}

\section{Results}
\label{R}
In this section we present the results from our eight models using \sc{primal}\rm. We will first show the results for the smooth featureless target disc previously explored in Paper 1. Then we apply \sc{primal }\rm to three different barred targets. We also examine how \sc{primal }\rm can reproduce the target galaxy with a partial data set of the observables, or an incorrect bar angle, or without the rotating reference frame using one of the targets. Table \ref{Mpars} shows which target the model is recreating, the bar pattern speeds, the likelihood values for radial, tangential and rotation velocity, $\mathcal{L}$, in equation (\ref{L}) and the $\chi^2_{\rho}$ for the density, where
\begin{equation}
\chi^{2}_{\rho} = \frac{\sum \Delta_{\rho}^{2}}{N_r}.
\label{chi2}
\end{equation}
Note that we include only particles within 10 kpc, and $N_r$ is the number of particles satisfying this criteria. Note also that in the likelihood case, although we seek to maximise likelihood, the values are $-\mathcal{L}$, and hence smaller values in Table \ref{Mpars} mean higher likelihood. The absolute values of $\mathcal{L}$ are not important. We cannot compare these values between models for different targets. However, the relative differences in $\chi^2_{\rho}$ and in $\mathcal{L}$ between models for the same target observables are meaningful and are used in Section 3.4.

\subsection{Smooth disc}
First, we demonstrate that the newly introduced likelihood velocity constraints can reproduce the smooth featureless disc target used in Paper 1. Model A applies \sc{primal }\rm to Target I, which was the target used in Paper 1, but using a larger number of particles. Note the high value of $\sigma_r^2/\sigma_z^2$ of Target I in Table 1 was used to deliberately suppress structure formation. Fig. \ref{AG} shows that the radial profiles of the density, the radial velocity dispersion, the vertical velocity dispersion and the mean rotation velocity for the target galaxy, the initial galaxy and the final output after \sc{primal }\rm is applied. The figure demonstrates that \sc{primal }\rm with the likelihood-based velocity constraints equally or even more accurately reproduces the target galaxies compared to our old version of the particle-by-particle M2M  (Paper 1). However, a quantitative comparison between the old and new version is not the main focus of this paper. As discussed above, we introduced the likelihood-based velocity constraints, because we can compare the velocity more directly and also introduce different errors for individual velocity components and individual particles. Therefore, the likelihood-based velocity constraints are a necessary update, and a comparison with the old version is not an important issue. Note that the properties shown in Fig. \ref{AG} are not explicitly constrained by \sc{primal}\rm. As discussed previously in Paper 1, it is interesting to note that although our particle-by-particle M2M uses only the first moment of the velocity components as observables, because \sc{primal }\rm tries to reproduce the velocity of individual particles, the velocity distribution becomes close to the target and therefore the velocity dispersion can be reproduced as well.

\subsection{Barred disc galaxies}
In this section we present the results of Models B, C and D, where we apply the same parameter set for \sc{primal }\rm to model three different target barred galaxies. Target II is a barred disc galaxy showing faint spiral structure. Fig. \ref{AO} shows the face-on and edge-on views of Target II and the final state of Model B (two top panels). The final model reproduces the bar feature very well. The observables are only constrained within 10 kpc of the Galactocentric radius and hence the areas outside this radius are reproduced with less accuracy. 

Fig. \ref{BG} shows the radial profiles of the surface density, radial and tangential velocity dispersion, and the mean rotation velocity for the target and the final model compared to the initial model. As in Paper 1 and Model A, these radial profiles are not directly constrained by \sc{primal}\rm, but are reproduced remarkably well. Fig. \ref{BG} shows a substantial increase in radial velocity dispersion and a corresponding decrease in mean rotational velocity from the initial to the final model. We believe that this is due to heating from the bar which leads to an excellent agreement with the velocity dispersion of the target. 



The pattern speed of the bar, $\Omega_p$, is also reproduced very well, as shown in Tables \ref{Tpars} and \ref{Mpars}. We calculate the pattern speed of the bar, $\Omega_p$, by calculating the change in angle of the bar between timesteps, divided by the difference in time between the steps. We take $\Omega_p$ to be the mean value from the final ten steps. We found that the bar pattern speed of the model is $\Omega_p=$ 27.9 km s$^{-1}$ which is close to that of the target, $\Omega_{t,p}=$ 27.5 km s$^{-1}$. This is probably due to our self-consistent calculation of the gravitational potential, because once the mass distribution and kinematic properties of the target disc are reproduced, a bar with a similar shape and pattern speed to those of the target is expected to develop. This is certainly helped by our use of a known, fixed dark matter halo potential. We are pleased to see a spiral arm developing in the model, which looks similar to the one seen in the target.

Model C applies \sc{primal }\rm to Target III, which is also a barred disc galaxy, but with a smaller bar than Target II, and a boxy and peanut shaped bulge \citep[e.g.][]{P84,AM02,DCMM05,BAADBF06,SZMMGH11}, as can be seen in the top panel of Fig. \ref{CTM}. Rather surprisingly \sc{primal }\rm reproduces the boxy structure of the target as shown in the bottom panel of Fig. \ref{CTM}. Fig. \ref{CG} shows the radial profiles for Model C. We see a slight inaccuracy in the inner 2 kpc of the radial velocity dispersion, and also in the rotational velocity in the inner 4 kpc, which corresponds roughly with the length of the bar. In addition, $\sigma_z$ is systematically higher than the target at all radii. As such the bar pattern speed is not as well reproduced as with Model B, with $\Omega_p=$ 30.4 km s$^{-1}$ compared to $\Omega_{t,p}=$ 31.7 km s$^{-1}$. However, we still think that this is a reasonably good recovery of the target, and it is encouraging for further developments to apply \sc{primal }\rm to more complicated observational data.


Model D takes Target IV which is morphologically similar to Target III, with a small bar and boxy peanut feature, as can be seen in Fig. \ref{DTM}. We see a slightly larger bar in the model than in the target. Fig. \ref{DG} shows slight inaccuracies in the recovery of the radial and vertical velocity dispersion and mean rotational velocity in the inner 3 kpc roughly consistent with the radius of the bar. However the pattern speed is still recovered well with $\Omega_p=$ 28.3 km s$^{-1}$ for the final model compared to the target of $\Omega_{t,p}=$ 28.9 km s$^{-1}$.

\subsection{Working with partial data}
Even with the huge amount of data returned by Gaia and related stellar surveys, due to our position within the Milky Way's disc, we will not even come close to having a complete data set of the disc stars. Therefore it is important to make sure our method is still applicable when we do not have access to the complete picture of the disc. Our previous models have used all data within 10 kpc from the galactic centre. However Model E was performed with a simple selection function restricting the observable volume to a 10 kpc sphere around a point in the plane 8 kpc from the galactic centre, i.e. at $(x,y,z)=(8,0,0)$ in Fig. \ref{AO}, roughly emulating Gaia's observable area, while ignoring effects such as extinction and errors. This is merely the first step towards using \sc{primal }\rm with realistic data.
\begin{figure}
\centering
\resizebox{\hsize}{!}{\includegraphics{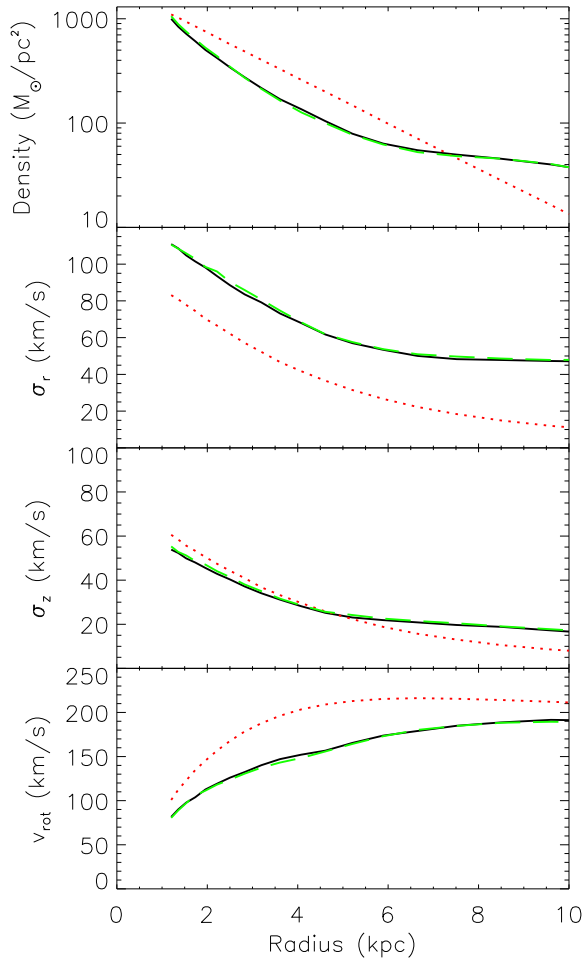}}
\caption{Same as Fig. \ref{AG}, but for Model E with Target II, performed with a partial data set.}
\label{EG}
\end{figure}

The third panel of Fig. \ref{AO} shows the face-on and edge-on view of Model E, which has a similar bar to the target (top panel), with a hint of a spiral arm in the lower left quadrant matching the one visible in the target. Fig. \ref{EG} shows that an excellent agreement of the final model with the target radial profiles is still obtained with the restricted data set. This is an improvement on Paper 1 where we saw loss of accuracy when the observable field was restricted. We believe that this is helped by both the likelihood form of velocity observable and the higher resolution with which the simulations have been carried out. The bar pattern speed is recovered very well with $\Omega_p=$ 27.3 km s$^{-1}$ compared to the target of $\Omega_{t,p}=$ 27.5 km s$^{-1}$. This shows the ability of \sc{primal }\rm to produce reasonable results when supplied with a partial data set of the disc particles. However we are aware that this selection function is crude and the next stage of our work will deal with more realistic selection functions and expected observational errors. 

\begin{figure}
\centering
\resizebox{\hsize}{!}{\includegraphics{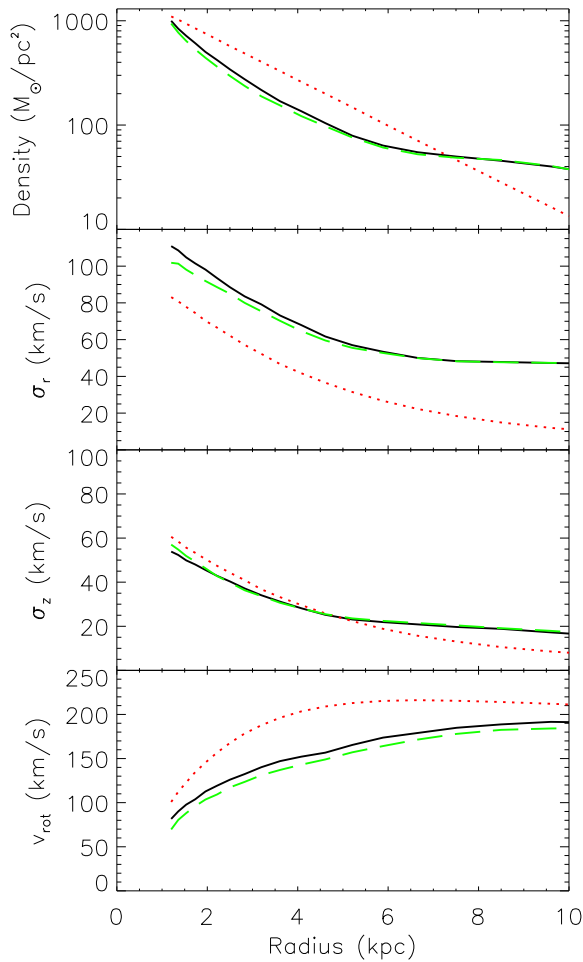}}
\caption{Same as Fig. \ref{AG}, but for Model F with Target II, performed with the assumed angle of the bar offset from 30$^{\text{o}}$ compared to the real value.}
\label{FG}
\end{figure}
\begin{figure}
\centering
\resizebox{\hsize}{!}{\includegraphics{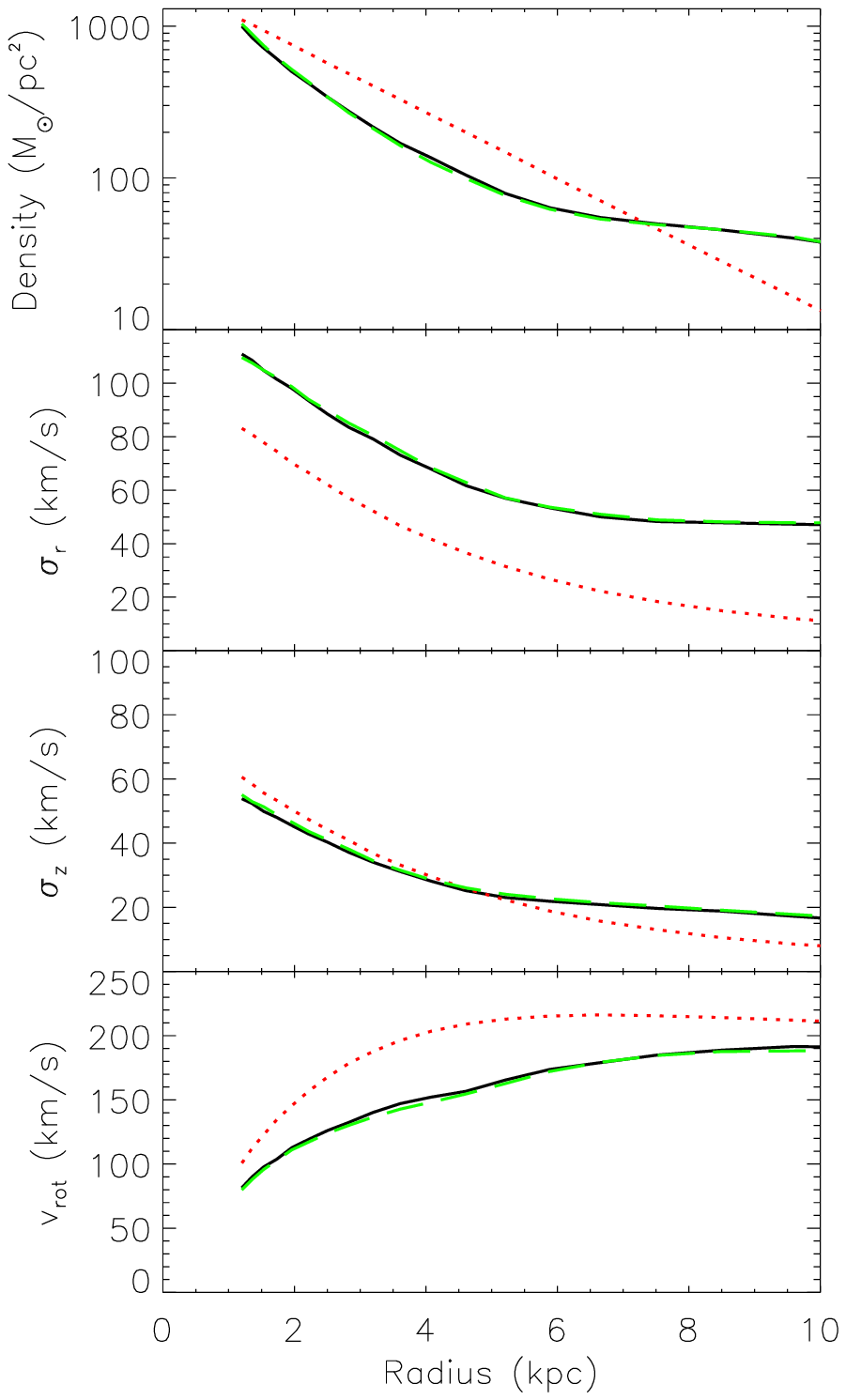}}
\caption{Same as Fig. \ref{AG}, but for Model G with Target II, performed with the assumed angle of the bar offset from 10$^{\text{o}}$ compared to the real value.}
\label{GG}
\end{figure}
\begin{figure}
\centering
\resizebox{\hsize}{!}{\includegraphics{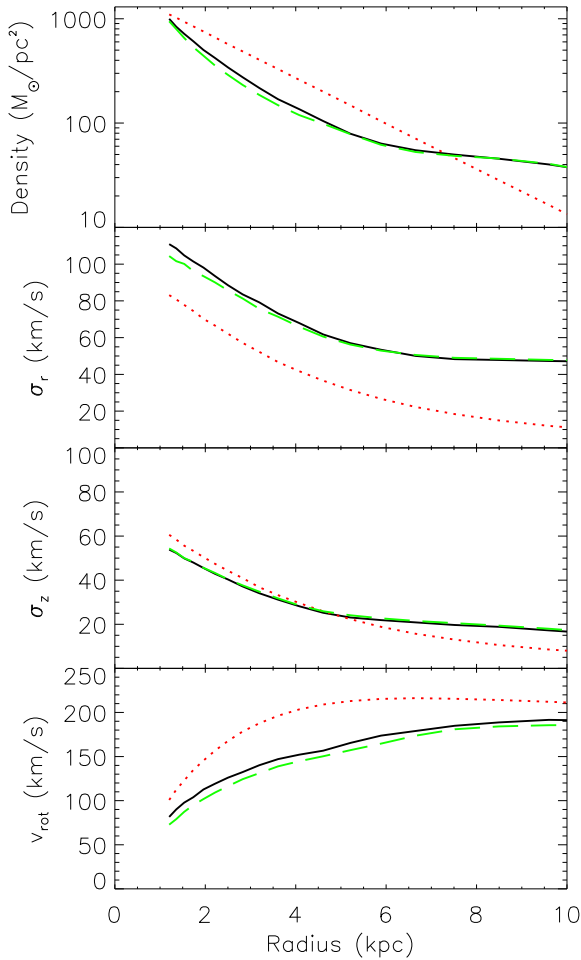}}
\caption{Same as Fig. \ref{AG}, but for Model H with Target II, which is performed without the rotating reference frame.}
\label{HG}
\end{figure}

\subsection{Working with an incorrect bar angle}
As mentioned in Section \ref{intro-sec}, the bar angle of the Milky Way is still still debated. Ultimately, we aim to recover the dynamical state of the Milky Way with \sc{primal }\rm from the future stellar survey data, and recovering the bar angle is also one of our targets. In the previous sections, we assumed that the bar angle of the target is known and we align the bar of the model galaxy to that of the target at every timestep to evaluate the observables. If we do not know the bar angle of the target, like with the Milky Way, we could try different bar angles and hope that the models with the lowest $\chi^2$ and/or the maximum likelihood values recover the bar angle of the target, which is the strategy taken by \cite{LMIII}. In this section, we examine the effects of running \sc{primal }\rm with an incorrect bar angle. Models F and G are performed with the bar angles deliberately set to be incorrect by 30 and 10 degrees respectively. In this section we again use all data within 10 kpc from the galactic centre.

Model F has been performed while assuming that the bar angle is $30^{\text{o}}$ less than the real angle of the target. The fourth panel of Fig. \ref{AO} shows a poor reproduction of the target bar morphology in Model F, which is significantly shorter than that of the target (top panel). We also see no evidence of the spiral structure seen in other cases. Fig. \ref{GG} shows the radial profiles for Model F. There is a discrepancy in the inner 4 kpc of the model compared to the target in both the density profile and the radial velocity dispersion. This is in agreement with the weaker bar shown in Fig. \ref{AO}. The average rotational velocity is also lower across the disc. This is also reflected in the final pattern speed of $\Omega_p=$ 23.6 km s$^{-1}$ compared to the target of $\Omega_{t,p}=$ 27.5 km s$^{-1}$. However, in the real Milky Way case, we cannot know the correct profiles or the bar pattern speed in advance. On the other hand, we can evaluate the goodness of fit by $\chi^2_{\rho}$ or the values of the likelihood, $\mathcal{L}_v$. In Table \ref{Mpars}, Model F shows significantly worse values of $\chi^2_{\rho}$ and $\mathcal{L}_v$ than those of Model B which assumes the correct bar angle. Therefore, we should be able to tell easily if the bar angle is off by 30 degrees, at least in this simple target case.

Model G has been performed while assuming that the bar angle is 10 degrees less than the bar angle of the target. The fifth panel of Fig. \ref{AO} shows a barred disc which is morphologically similar to the target (top panel). The bar is reproduced well whereas the spiral structure is barely visible. Fig. \ref{GG} shows the radial profiles for Model G, which again reproduces very well those of the target. The bar pattern speed is still well recovered with $\Omega_p=$ 28.0 km s$^{-1}$ compared to the target of $\Omega_{t,p}=$ 27.5 km s$^{-1}$. In Table \ref{Mpars}, Model G shows similar values of $\chi^2_{\rho}$ and $\mathcal{L}_v$ to those of Model B, although the velocity likelihood values are slightly worse. These results may indicate that \sc{primal }\rm does not have the power to determine the bar angle within 10 degree accuracy, but can recover it with better than $30^{\text{o}}$ accuracy. However, our ultimate target is much more complicated than this ideal target, and at this stage we do not explore further the expected accuracy of recovering the correct bar angle for this ideal target. At least we demonstrate that with this type of exercise we can examine how accurately the dynamical model, such as \sc{primal}\rm, can recover the bar angle. In our future study, we will construct more realistic mock observational data from $N$-body barred simulated discs and `train' \sc{primal }\rm to recover the bar angle as accurately as possible, and finally evaluate the expected accuracy of our recovered bar angle using the comparison demonstrated in this section.

\subsection{The importance of the rotating reference frame}
\label{NRF}
In this section we show a brief comparison between the resulting models with and without the rotating reference frame. Model H was performed under identical conditions to Model B, but without the reference frame corrections detailed in Section \ref{RF}. The bottom panel of Fig. \ref{AO} shows that the resulting disc contains a less prominent bar, and no evidence of spiral structure in a similar fashion to Model F. Fig. \ref{HG} shows the radial profiles of Model H. In the inner 4 kpc region, the radial density and radial velocity profiles are lower for the model than for the target. The average rotation velocity is lower than that of the target across the whole disc. The pattern speed is also too low with $\Omega_p=$ 24.3 km s$^{-1}$ compared to the target of $\Omega_{t,p}=$ 27.5 km s$^{-1}$. 
\begin{figure}
\centering
\resizebox{\hsize}{!}{\includegraphics{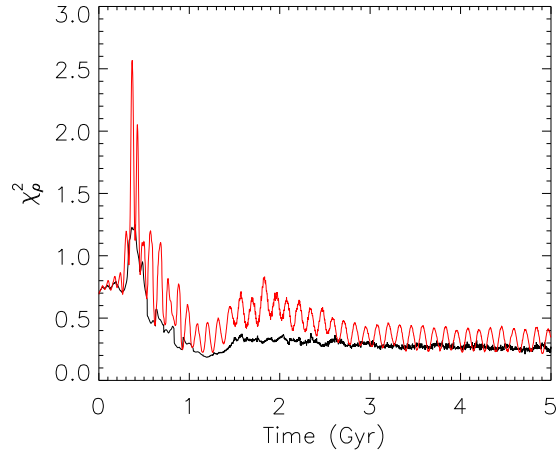}}
\caption{Time evolution of $\chi^2_{\rho}$ for density for Model B (black line) compared to Model H (red line).}
\label{Xi}
\end{figure}
Fig. \ref{Xi} shows a comparison of the evolution of the $\chi^2_{\rho}$ of the density between Model B and Model H. The $\chi^2_{\rho}$ in Model H experiences periodic oscillations in time with the bar rotation which are not seen in Model B. This lack of a smooth model convergence along with the poor accuracy on the recovered radial profiles shows the importance of having a rotating reference frame.

\section{Summary}
\label{SF}
We have demonstrated that our updated particle-by-particle M2M algorithm, \sc{primal}\rm, can recover a target disc system with a bar, including boxy/peanut features, in a known dark matter halo potential. In \sc{primal}\rm, the observables are compared with the model at the position of the target particles. The mass of the model particles are adjusted to reproduce the target observables, and the gravitational potential is calculated self-consistently from the model particle mass distribution. We have introduced the likelihood-based velocity constraints to \sc{primal}\rm, which allows us to compare the velocity of the target particle more directly than the smoothed velocity field used in our previous algorithm. To apply this method to a barred disc, we evaluate at every timestep the density and velocity likelihood after the reference frame of the model disc has been corrected, so that the bar of the model is always aligned with the bar of the target. Our fiducial model recovers the radial profiles of the surface density, the radial and vertical velocity dispersion and the mean rotation velocity of the target system very well. In addition, because of our self-gravity implementation of M2M, we can reproduce the bar morphology and pattern speed. We have demonstrated that \sc{primal }\rm performs well even when the observables are restricted to within a sphere of radius 10 kpc around a point in the disc plane 8 kpc from the centre.

We admit that these applications are still simplified cases. The excellent recovery of the properties of the target galaxies is not surprising because we have not applied any observational errors or selection functions. Our ultimate goal is to further improve \sc{primal }\rm to be applicable to the future stellar survey data, including the Gaia data. While Gaia will return an unprecedentedly large amount of data, for approximately one billion stars, the accuracy of this data will be highly variable due to distance, extinction, location in the sky etc. In a forthcoming paper, we will apply \sc{primal }\rm to more realistic mock observational data from $N$-body simulations, taking into account the observational errors and selection functions. This paper also assumes a known, static, spherical dark matter halo potential for simplicity. In reality the dark matter halo remains very much unknown and yet has a significant effect on the dynamics of its inner galaxy. Thus we need to explore different dark matter halo potentials and to consider the possibility of using a model which includes a live halo.



We remain optimistic that we can continue to improve \sc{primal}\rm, and develop a unique tool to recover the dynamical properties of the Milky Way from the future large-scale stellar survey data.

\section*{Acknowledgements}
We thank the anonymous referee for their time and effort. We thank Victor Debattista for recommending their likelihood-based velocity constraints. We gratefully acknowledge the support of the UK's Science \& Technology Facilities Council (STFC Grant ST/H00260X/1). The calculations for this paper were performed on Cray XT4 at Center for Computational Astrophysics, CfCA, of the National Astronomical Observatory of Japan and the DiRAC Facilities, Legion and COSMOS, jointly funded by the STFC and the Large Facilities Capital Fund of BIS. The authors also thank the support of the STFC-funded Miracle and COSMOS consortium (part of the DiRAC facility) in providing access to the UCL Legion and Cambridge COSMOS High Performance Computing Facilities. We additionally acknowledge the support of UCL's Research Computing team with the use of the Legion facility.  This work was carried out, in part, through the Gaia Research for European Astronomy Training (GREAT-ITN) network. The research leading to these results has received funding from the European Union Seventh Framework Programme ([FP7/2007-2013] under grant agreement number 264895). HM is supported by the Natural Sciences and Engineering Research Council of Canada, and the Canada Research Chair program. HM is thankful to the Department of Physics and Astronomy, University of Victoria, for its hospitality.

\bibliographystyle{mn2e}
\bibliography{ref2}

\label{lastpage}
\end{document}